\documentclass[twocolumn]{aastex7}

\usepackage{amsmath}
\usepackage{wasysym}
\usepackage{savesym}
\savesymbol{tablenum}
\usepackage{siunitx}
\restoresymbol{SIX}{tablenum}
\usepackage{booktabs}
\usepackage{rotating}
\usepackage[T1]{fontenc}

\DeclareSIUnit\Mearth{M_\oplus}
\DeclareSIUnit\Mmercury{M_M}
\DeclareSIUnit\Mjupiter{M_J}
\DeclareSIUnit\Rearth{R_\oplus}
\DeclareSIUnit\Rjupiter{R_J}
\DeclareSIUnit\Mmoon{M_\textrm{\leftmoon}}
\DeclareSIUnit\JEM{J_{EM}}
\DeclareSIUnit{\year}{yr}
\DeclareSIUnit\Gal{Gal}

\graphicspath{{./}{figures/}}

\shorttitle{The Caloris basin-forming impact}
\shortauthors{Meier et al.}
\received{27.08.2026}
\submitjournal{ApJ}

\begin{document}

\title{Three-dimensional SPH simulations of the Caloris basin-forming impact: basin scaling, the gravity anomaly, and antipodal effects}

\author[orcid=0000-0001-9682-8563,gname=Thomas, sname=Meier]{Thomas Meier} 
\affiliation{Department of Astrophysics, University of Zurich, Winterthurerstrasse 190, CH-8057 Zurich, Switzerland}
\email[show]{thomas.meier5@uzh.ch}

\author[orcid=0000-0002-4535-3956,gname=Christian, sname=Reinhardt]{Christian Reinhardt}
\affiliation{Department of Astrophysics, University of Zurich, Winterthurerstrasse 190, CH-8057 Zurich, Switzerland}
\affiliation{Physics Institute, Space Research and Planetary Sciences, University of Bern, Sidlerstrasse 5, CH-3012 Bern, Switzerland}
\email{christian.reinhardt@uzh.ch}

\author[orcid=0000-0002-1800-2974,gname=Martin, sname=Jutzi]{Martin Jutzi} 
\affiliation{Physics Institute, Space Research and Planetary Sciences, University of Bern, Sidlerstrasse 5, CH-3012 Bern, Switzerland}
\email{martin.jutzi@unibe.ch}

\author[orcid=0000-0001-7565-8622,gname=Joachim, sname=Stadel]{Joachim Stadel} 
\affiliation{Department of Astrophysics, University of Zurich, Winterthurerstrasse 190, CH-8057 Zurich, Switzerland}
\email{joachimgerhard.stadel@uzh.ch}

\begin{abstract}
The Caloris basin is the largest well-preserved impact structure on Mercury, yet its formation conditions, the origin of its positive gravity anomaly, and its relation to the antipodal terrain remain poorly constrained. We present global, three-dimensional smoothed particle hydrodynamics simulations of the Caloris basin-forming impact with the \texttt{pkdgrav3} code, including material strength. We survey 225 combinations of impactor radius, velocity, angle, and target thermal profile, complemented by simulations of up to two billion particles, which resolve the adopted \SI{40}{\kilo\meter} crust by ten particle layers. At all resolutions, basin sizes are measured directly from the crust particles. The measured basin diameters follow a single power law in impactor radius, velocity, and angle, systematically steeper than idealized point-source crater scaling. The observed Caloris diameter is reproduced by a broad family of impactors favoring oblique incidence; sensitivity to resolution and crust thickness shifts this family toward smaller or slower impactors. All impactor material remaining in the basin region is vaporized, so our crust-like impactors leave no buried remnant. Instead, the impact thins the mantle and raises a local dome on Mercury's core. The dome's isolated gravity signal is positive and centered on the basin, supporting the mantle-uplift origin proposed for the observed mascon. At the antipode, single seismic pulses cannot loft surface material when strength is included, yet accumulated strain exceeds the elastic limit. A kilometer-scale equivalent thickness of impact-derived material then converges on this weakened surface. Ejecta convergence contributed at least as much as seismic shaking to forming the terrain.
\end{abstract}



\section{Introduction}\label{sec:Introduction}
Mercury is the smallest and innermost planet of the Solar System and possesses several properties that distinguish it from the other terrestrial planets. Despite its modest size, its mean density of \SI{5.43}{\gram\per\centi\meter\tothe3} is comparable to that of Earth, implying an unusually large metallic core that comprises roughly \SI{70}{\percent} of the planet's mass \citep{spohnInteriorStructureMercury2001,hauckiiCuriousCaseMercurys2013}. In contrast to Earth, Mercury exhibits neither plate tectonics nor a substantial atmosphere, allowing impact structures formed early in Solar System history to remain remarkably well preserved \citep{wattersMercurysCrustalThickness2021}. As a consequence, Mercury provides a unique laboratory for investigating large planetary impacts and their long-term geophysical consequences.

The most prominent impact structure on Mercury is the Caloris basin, a multiring basin exceeding \SI{1500}{\kilo\meter} in diameter that formed during the Late Heavy Bombardment approximately \SIrange{3.9}{3.7}{\giga\year} ago \citep{murchieGeologyCalorisBasin2008,rodriguezChaoticTerrainsMercury2020}. The basin was first identified during the Mariner 10 mission, although only a fraction of its extent was imaged \citep{murrayMariner10Pictures1974}. Mariner 10 also revealed a region of hilly and lineated terrain located approximately antipodal to Caloris, leading to the hypothesis that seismic waves generated by the impact were focused on the opposite side of the planet and produced the observed terrain \citep{schultzSeismicEffectsMajor1975,hughesGlobalSeismicEffects1977,luSeismicEffectsCaloris2011}. More than three decades later, the MESSENGER mission \citep{solomonMESSENGERMissionMercury2001} mapped the basin in its entirety and provided global topography, geochemical measurements, and high-resolution gravity data, revealing a pronounced positive gravity anomaly spatially correlated with the basin \citep{mazaricoGravityFieldOrientation2014,smithGravityFieldInternal2012}. Together with crustal thickness estimates, these observations suggest that the Caloris-forming impact substantially modified Mercury's crust and mantle, potentially through crustal thinning, emplacement of dense material into the mantle, and mantle melting that subsequently drove extensive volcanism surrounding the basin \citep{robertsEffectCalorisImpact2012,johnsonWholeNewMercury2016}.

Despite this wealth of observations, the impact conditions that produced the Caloris basin remain poorly constrained. Published estimates of the impactor size span diameters of $\sim\SIrange{80}{525}{\kilo\meter}$, so the inferred impact energies differ by more than an order of magnitude. The smaller values come from the best-fit bodies of axisymmetric hydrocode models \citep{ivanovBasinformingImpactsReconnaissance2010,potterBasinFormationMercury2017,gosselinCrustalBlockMuted2023} and from the projectiles of interior-response models, which prescribe the impact heating rather than simulating the collision \citep{robertsEffectCalorisImpact2012}; the much larger upper value is suggested by an analysis of the basin-related sculpture \citep{schultzSizesNatureBasin2017}. Moreover, the hydrocode studies to date have relied on axisymmetric geometries, which are necessarily restricted to head-on impacts and cannot capture the inherently three-dimensional dynamics of oblique collisions, asymmetric excavation, or the global propagation of shock waves responsible for antipodal effects. The influence of impact angle and realistic impact geometries on the formation of the Caloris basin has thus remained largely unexplored.

In this work, we investigate the formation of the Caloris basin using global three-dimensional smoothed particle hydrodynamics (SPH) simulations performed with the \texttt{pkdgrav3} code. We explore a broad parameter space of impact velocity, angle, and impactor size to identify collisions capable of reproducing the observed basin morphology. The highest-resolution simulation contains two billion particles, corresponding to a mean inter-particle spacing of $\approx\SI{3}{\kilo\meter}$, to our knowledge the highest resolution yet applied to a global, three-dimensional simulation of a basin-forming impact; axisymmetric two-dimensional models reach kilometer-scale resolution, but at the price of the geometric restrictions discussed above. This enables us to resolve the excavation process, mantle deformation, and redistribution of impactor material in detail. At this resolution, Mercury's crust is resolved by ten particle layers, allowing the crustal response to the impact to be captured directly rather than treated as a poorly resolved interface. For this highest-resolution simulation, we adopt a crustal thickness of \SI{40}{\kilo\meter}, at the upper end of current observational estimates; the survey simulations use thicker crusts as a numerical necessity (Section~\ref{sec:Models}). By comparing the simulated basin structure, crustal modification, and mantle density anomalies with observational constraints from MESSENGER, we investigate the origin of the Caloris gravity anomaly. For the vertical-incidence case, we also quantify the impact's antipodal effects: the focused seismic shaking and the convergence of impact-derived material. To our knowledge, this is the first study to address the basin size, the gravity anomaly, and the antipodal effects of the Caloris-forming impact within a single, self-consistent simulation framework. These results provide new constraints on one of the largest impact events recorded in the inner Solar System and establish a framework for interpreting future observations from the BepiColombo mission \citep{benkhoffBepiColomboComprehensiveExploration2010,benkhoffBepiColomboMissionOverview2021}.

This paper is structured as follows. Section~\ref{sec:Methods} describes the SPH methodology employed for the impact simulations, the equations of state used, and the generation of the planetary models and initial conditions. Section~\ref{sec:Results_and_Discussion} presents and discusses the results of the simulations. Finally, Section~\ref{sec:Summary_and_Conclusions} summarizes our findings and provides our conclusions.

\section{Methods}\label{sec:Methods}
We model the collisions using the Smoothed Particle Hydrodynamics (SPH) method \citep{lucyNumericalApproachTesting1977,monaghanSmoothedParticleHydrodynamics1992,springelSmoothedParticleHydrodynamics2010}, which has been extensively applied to the study of giant impacts \citep[e.g.][]{canupSimulationsLateLunarforming2004,asphaugMercuryOtherIronrich2014,emsenhuberSPHCalculationsMarsscale2018,reinhardtFormingIronrichPlanets2022,kegerreisImmediateOriginMoon2022,timpeSystematicSurveyMoonforming2023,ballantyneInvestigatingFeasibilityImpactinduced2023,meierSystematicSurveyMoonforming2024,ballantyneSputnikPlanitiaImpactor2024}. The simulations are performed with the \texttt{pkdgrav3} code \citep{potterPKDGRAV3TrillionParticle2017,meierSmoothedParticleHydrodynamics2026,meierSmoothedParticleHydrodynamics2026a}, which implements a modern SPH formulation based on \citet{springelCosmologicalSmoothedParticle2002,priceSmoothedParticleHydrodynamics2012}. This formulation supports adaptive smoothing lengths and employs Wendland kernels, which avoid the pairing instability and allow accurate density estimates with large neighbor numbers of up to 400 particles \citep{dehnenImprovingConvergenceSmoothed2012}. All simulations in this work use the Wendland~C6 kernel with a target number of 400 neighbors.

The code incorporates several recent developments specifically tailored for giant impact simulations, including corrections for interfaces and free surfaces \citep{reinhardtNumericalAspectsGiant2017,reinhardtBifurcationHistoryUranus2020,ruiz-bonillaDealingDensityDiscontinuities2022}, an entropy-conserving energy formulation \citep{reinhardtNumericalAspectsGiant2017}, and a generalized equation-of-state (EOS) interface \citep{meierEOSResolutionConspiracy2021,meierEOSlib2021}, enabling simulations involving multiple materials with distinct EOS models. \texttt{pkdgrav3} has previously been employed for a wide range of impact scenarios, including planetary collisions and satellite-forming impacts \citep{matzkevichOutcomeCollisionsGaseous2024,meierSystematicSurveyMoonforming2024,meierOriginJupitersFuzzy2025,bussmannPossibilityGiantImpact2025}. The code is highly scalable on modern high-performance computing (HPC) architectures, supporting both CPU-only and hybrid CPU/GPU configurations and enabling simulations with several billion particles.

\subsection{Equations of state and material strength}\label{sec:EOS}
\begin{table*}[ht!]
\centering
\begin{tabular}{llccc}
\toprule
Parameter & Description & Iron & Forsterite & Quartz \\
\midrule
$\Gamma$ & Shear modulus & \SI{76}{\giga\pascal} & \SI{72}{\giga\pascal} & \SI{72}{\giga\pascal} \\
$Y_{\rm m}$    & Shear strength at $P\rightarrow\infty$ & \SI{0.68}{\giga\pascal} & \SI{3.5}{\giga\pascal} & \SI{3.5}{\giga\pascal} \\
$Y_0$    & Cohesion & \num{0} & \num{0} & \num{0} \\
$\mu_{\rm i}$  & Coefficient of friction (intact material) & \num{0.8} & \num{0.8} & \num{0.8} \\
$\mu_{\rm d}$  & Coefficient of friction (damaged material) & \num{0.8} & \num{0.8} & \num{0.8} \\
$\xi$    & Thermal softening parameter & \num{1.2} & \num{1.2} & \num{1.2} \\
\bottomrule
\end{tabular}
\caption{Parameters of the strength model \citep{collinsModelingDamageDeformation2004,meierSmoothedParticleHydrodynamics2026a}. With zero cohesion and $\mu_{\rm i}=\mu_{\rm d}$, the model reduces to the single saturating yield envelope of Equation~\eqref{eq:yield}, describing fully damaged material (Section~\ref{sec:EOS}); the two silicate materials share a single parameter set representative of competent rock.}
\label{tab:Strength_parameters}
\end{table*}

To model the materials involved in the simulations, we use the equations of state (EOSs) for iron \citep{stewartEquationStateModel2020a}, forsterite \citep{stewartEquationStateModel2019} and quartz \citep{meloshHydrocodeEquationState2007} constructed from M-ANEOS \citep{thompsonImprovementsCHARTRadiationhydrodynamic1974, meloshHydrocodeEquationState2007, thompsonMANEOS2019}. Following \citet{jutziSPHCalculationsAsteroid2015,emsenhuberSPHCalculationsMarsscale2018}, material strength is modeled using an elastic--plastic formulation  \citep{benzImpactSimulationsFracture1994,benzSimulationsBrittleSolids1995} with a pressure-dependent shear strength described by the yield envelope of \citet{collinsModelingDamageDeformation2004} and thermal softening following \citet{ohnakaShearFailureStrength1995}, as implemented in \citet{meierSmoothedParticleHydrodynamics2026a}. At the scale of a basin-forming impact, the target lithosphere is expected to be pervasively fractured, so we describe the target as fully damaged, cohesionless material \citep{collinsModelingDamageDeformation2004}. Its shear strength is frictional at low pressure and cannot exceed a limiting strength at high pressure; both properties are captured by the smoothly saturating yield envelope

\begin{equation}
Y(P) = \frac{\mu_{\rm d} P}{1 + \mu_{\rm d} P / Y_{\rm m}}\,, \label{eq:yield}
\end{equation}

\noindent where $\mu_{\rm d}$ is the friction coefficient of damaged material and $Y_{\rm m}$ the limiting strength (Table~\ref{tab:Strength_parameters}). Within the general model of \citet{collinsModelingDamageDeformation2004}, this corresponds to the intact-material envelope with zero cohesion and $\mu_{\rm i} = \mu_{\rm d}$; the linear damaged branch $\mu_{\rm d} P$ then lies above Equation~\eqref{eq:yield} at all pressures and is never selected. This is a good approximation for intermediate-mass bodies \citep{jutziSPHCalculationsAsteroid2015,emsenhuberSPHCalculationsMarsscale2018,ballantyneInvestigatingFeasibilityImpactinduced2023,ballantyneSputnikPlanitiaImpactor2024,dentonCaptureAncientCharon2025}. Thermal softening is applied as the temperature approaches the melting curve. The corresponding parameters are listed in Table~\ref{tab:Strength_parameters}; for simplicity, the two silicate materials share a single, identical parameter set representative of competent rock, i.e., solid, consolidated rock as opposed to porous or unconsolidated material. Material strength is enabled in all simulations presented in this work except the single strengthless run of Section~\ref{sec:antipode_accel}, which serves only as a limiting case. For iron and forsterite, the melting curve, $T_{\rm melt}(\rho)$, is obtained directly from the M-ANEOS phase information. For quartz, we convert the $T_{\rm melt}(P)$ relation of \citet{gonzalez-cataldoMeltingCurveSiO22016} to $T_{\rm melt}(\rho)$ using the EOS.

\subsection{Models}\label{sec:Models}
\begin{figure}[ht!]
\centering
\includegraphics[width=\linewidth]{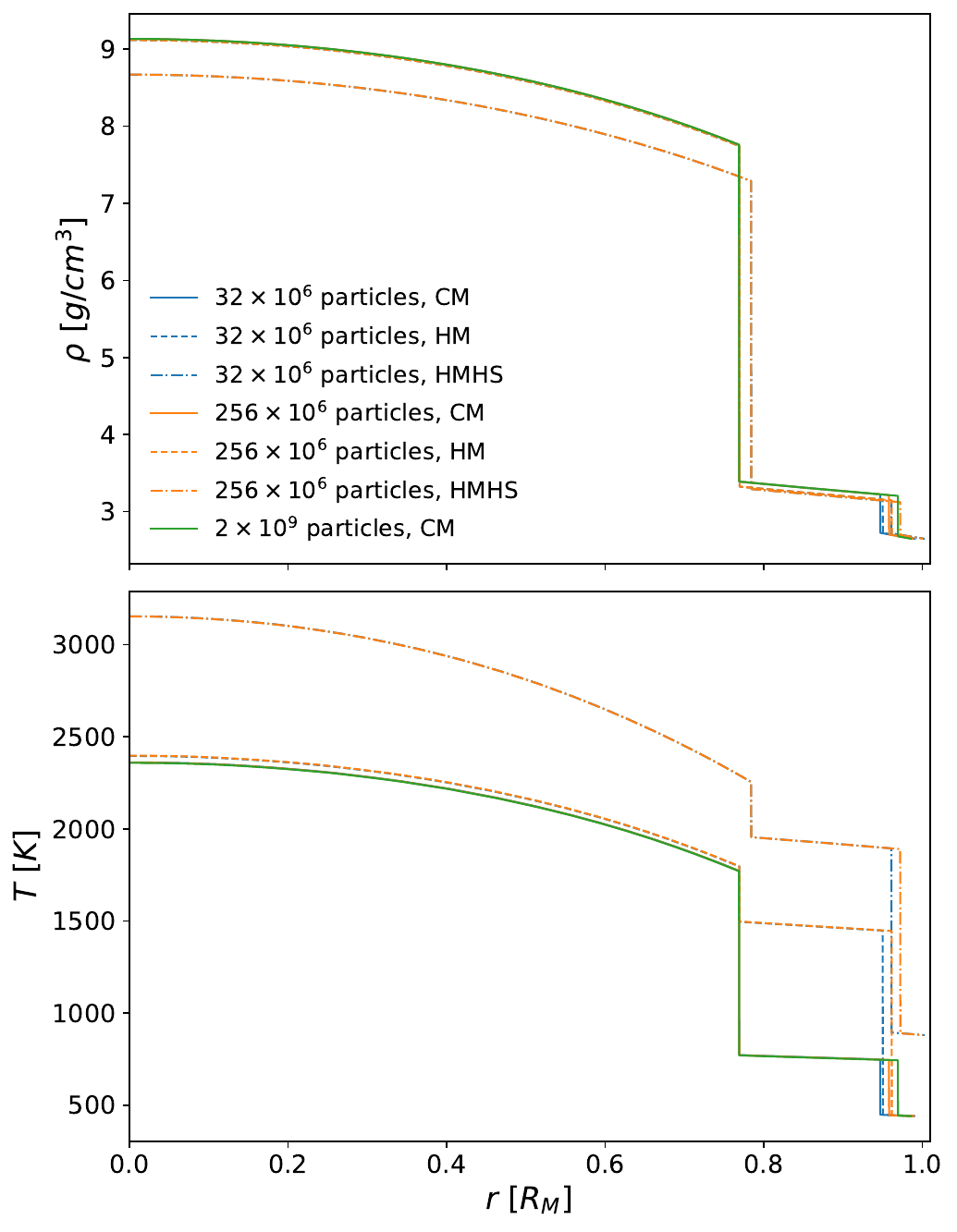}
\caption{Density (top) and temperature (bottom) profiles of the target initial states for the three thermal models: cold mantle (CM), hot mantle (HM), and hot mantle with hot surface (HMHS); see Section~\ref{sec:Models} for their definitions.}
\label{fig:Initial_profiles}
\end{figure}

Mercury is modeled as a differentiated body with a total mass of \SI{1}{\Mmercury} = \SI{0.055}{\Mearth}, consisting of an iron core containing \SI{70}{\percent} of the total mass, a forsterite mantle, and a quartz crust. Quartz serves here as a proxy for Mercury's crust, which is in reality a low-iron, Mg-rich silicate \citep[e.g.,][]{nittlerMajorelementCompositionMercurys2011}. We adopt it for two reasons: a well-calibrated M-ANEOS model is available (Section~\ref{sec:EOS}), and its reference density of \SI{2.65}{\gram\per\centi\meter\tothe3} gives a density contrast with the forsterite mantle comparable to the crust--mantle contrast on Mercury. This contrast is the property most relevant to the basin-size metric used in this work (Appendix~\ref{sec:appendix:Measuring_Crater_Size}). The true crustal density is likely somewhat higher \citep[$\approx\SIrange{2.7}{3.1}{\gram\per\centi\meter\tothe3}$;][]{soriThinDenseCrust2018}, which would proportionally reduce the equivalent deposit thicknesses quoted in Section~\ref{sec:antipode_mass}. The particle realizations are generated using the \texttt{ballic} code \citep{reinhardtNumericalAspectsGiant2017}, including extensions for multi-component planetary models \citep{chauFormingMercuryGiant2018,reinhardtBifurcationHistoryUranus2020}, which constructs hydrostatic, isentropic interior profiles for each material layer. The thermal state of Mercury's interior remains poorly constrained, particularly the mantle temperature profile \citep{michelThermalEvolutionMercury2013,tosiThermochemicalEvolutionMercurys2013}. We therefore consider two bracketing thermal structures, both with a core--mantle boundary (CMB) temperature of approximately \SI{1800}{\kelvin}. The hot-mantle model (HM) has a \SI{300}{\kelvin} temperature jump across the CMB, a \SI{1000}{\kelvin} jump across the mantle--crust boundary (MCB), and a surface temperature of \SI{440}{\kelvin}. The cold-mantle model (CM) instead has a \SI{1000}{\kelvin} jump across the CMB, a \SI{300}{\kelvin} jump across the MCB, and the same surface temperature. A third model (HMHS) is identical to the hot-mantle profile but has a surface temperature of \SI{880}{\kelvin}, representing a hotter interior state that may be characteristic of an earlier stage in Mercury's thermal evolution. The crust mass fraction is adjusted to ensure that the crust is resolved by several particle layers (7--11 layers) at each resolution, an empirical choice whose adequacy we assess directly with the series of runs varying resolution and crustal thickness in Section~\ref{sec:resolution_sensitivity}. We perform simulations at three nominal resolutions containing \SI{32e6}{}, \SI{256e6}{}, and \SI{2e9}{} particles, corresponding to mean inter-particle spacings of roughly \SI{12}{}, \SI{6}{}, and \SI{3}{\kilo\meter}. We quote the inter-particle spacing as the resolution measure; the SPH smoothing length, set by the \num{400} nearest neighbors used in this work, is $\approx2.3$ mean inter-particle spacings, so the formal spatial resolution is correspondingly coarser. The corresponding crust thicknesses are \SI{104}{\kilo\meter}, \SI{72.5}{\kilo\meter}, and \SI{40}{\kilo\meter}, respectively. Estimates of Mercury's mean crustal thickness span roughly \SIrange{26}{50}{\kilo\meter}, from the \SI{26\pm11}{\kilo\meter} of \citet{soriThinDenseCrust2018} and the \SI{35\pm18}{\kilo\meter} of \citet{padovanThicknessCrustMercury2015} to the \SIrange{23}{50}{\kilo\meter} range of \citet{konoplivMercuryGravityField2020}, with regionally resolved models yielding mean values of $\sim\SI{35}{\kilo\meter}$ over local variations of several tens of kilometers \citep{beutheMercurysCrustalThickness2020,wattersMercurysCrustalThickness2021}. The \SI{40}{\kilo\meter} crust of the highest-resolution model thus lies at the upper end of these observational estimates; this choice is additionally motivated by the requirement that the crust be resolved by ten particle layers at two billion particles. The thicker crusts of the lower-resolution models are a numerical necessity imposed by the same particle-layer requirement. In addition, we perform simulations with \SI{256e6}{} particles using the thicker crust (\SI{104}{\kilo\meter}) of the \SI{32e6}{}-particle models. The resulting density and temperature profiles of all initial models are shown in Figure~\ref{fig:Initial_profiles}.

The impactors are modeled as spherical, undifferentiated bodies composed entirely of quartz, with radii of \SI{100}{\kilo\meter}, \SI{150}{\kilo\meter}, and \SI{200}{\kilo\meter}. The impactor radii are chosen such that the impactor is resolved by an adequate number of particles at the lowest resolution; impactors comparable to the \SIrange{80}{150}{\kilo\meter}-diameter best-fit bodies of previous hydrocode studies \citep{potterBasinFormationMercury2017,gosselinCrustalBlockMuted2023} would be marginally resolved at 32 million particles. Their particle realizations are likewise generated using \texttt{ballic}, assuming an isentropic interior profile and a surface temperature of \SI{440}{\kelvin}. To maintain a consistent mass resolution between the colliding bodies, the number of impactor particles is adjusted such that the impactor particle mass matches that of the corresponding target model.

Both the target and impactor bodies are relaxed using a ramped velocity damper \citep[see Section~2.3.7 in][]{meierSmoothedParticleHydrodynamics2026} to reduce residual numerical noise and obtain stable initial configurations, minimizing artificial particle motions at the beginning of the simulations.

\subsection{Initial conditions}\label{sec:Initial_conditions}
Using the relaxed bodies, we construct the initial conditions by placing the target and impactor in the center-of-mass frame with an initial separation given by

\begin{align}
\Delta x &= 1.1 (R_{\rm tar} + R_{\rm imp})\,,\\
\Delta y &= (R_{\rm tar} + R_{\rm imp}) \sin{\theta}\,,
\end{align}

\noindent where $R_{\rm tar}$ and $R_{\rm imp}$ are the target and impactor radii, respectively, and $\theta$ is the impact angle. The relative impact velocity is applied in the $x$-direction and distributed between the target and impactor according to their masses, ensuring that the center of mass remains stationary.

At a resolution of \SI{32e6}{} particles, we perform a comprehensive parameter survey covering all combinations of target thermal structures and impactor sizes. For each configuration, we consider five impact velocities, $v_{\rm imp}\in\{12,21,30,39,48\}\,\si{\kilo\meter\per\second}$, and five impact angles, $\theta\in\{0,15,30,45,60\}\,\si{\degree}$, resulting in a total of 225 simulations. The velocity range is chosen to bracket the expected impact velocities at Mercury, which, owing to the planet's proximity to the Sun, are the highest of any terrestrial planet. Dynamical models of the impactor population predict mean impact velocities of roughly \SIrange{30}{43}{\kilo\meter\per\second} \citep{lefeuvreNonuniformCrateringTerrestrial2008,marchiNEWCHRONOLOGYMOON2009,lefeuvreNonuniformCrateringMoon2011}, with a broad underlying velocity distribution extending well below and above these mean values. This velocity range coincides with that adopted by \citet{robertsEffectCalorisImpact2012} for their study of the impact's interior response.

At a resolution of \SI{256e6}{} particles, we perform a more limited set of simulations. These include the cases with $R_{\rm imp}=\SI{150}{\kilo\meter}$, $v_{\rm imp}=\SI{21}{\kilo\meter\per\second}$, and $\theta=\SI{0}{\degree}$ for all three thermal profiles, as well as an additional set using the thicker crust model. In addition, the simulation with the CM profile and \SI{72.5}{\kilo\meter} crust is repeated with material strength disabled; this run is identical to its strength-enabled counterpart in every other respect and serves only as a limiting case for the antipodal analysis (Section~\ref{sec:antipode_accel}). Finally, the strength-enabled CM case is repeated once more with a per-particle accumulated-strain diagnostic enabled (Section~\ref{sec:antipode_strain}); this run is likewise identical in every other respect.

At the highest resolution of \SI{2e9}{} particles, we perform a single simulation using the cold mantle profile, with $R_{\rm imp}=\SI{150}{\kilo\meter}$, $v_{\rm imp}=\SI{21}{\kilo\meter\per\second}$, and $\theta=\SI{0}{\degree}$.

The 32-million-particle survey simulations and the 256-million-particle simulation with the CM profile and \SI{72.5}{\kilo\meter} crust are evolved for \SI{3.54}{\hour} after impact; the remaining 256-million-particle simulations and the 2-billion-particle simulation are evolved for \SI{1.77}{\hour}. As shown in Appendix~\ref{sec:appendix:Measuring_Crater_Size}, the basin dimensions are stationary over this interval, so the differing durations do not affect any of the comparisons presented below.

\section{Results and Discussion}\label{sec:Results_and_Discussion}
Section~\ref{sec:Basin_size} quantifies the basin size across the parameter survey and its implications for the Caloris-forming impactor, Section~\ref{sec:gravity} addresses the origin of the Caloris gravity anomaly, and Section~\ref{sec:antipode} assesses the two classical impact-related mechanisms proposed for the antipodal terrain.

\subsection{Basin size}\label{sec:Basin_size}
\begin{figure*}[ht!]
\centering
\includegraphics[width=0.95\linewidth]{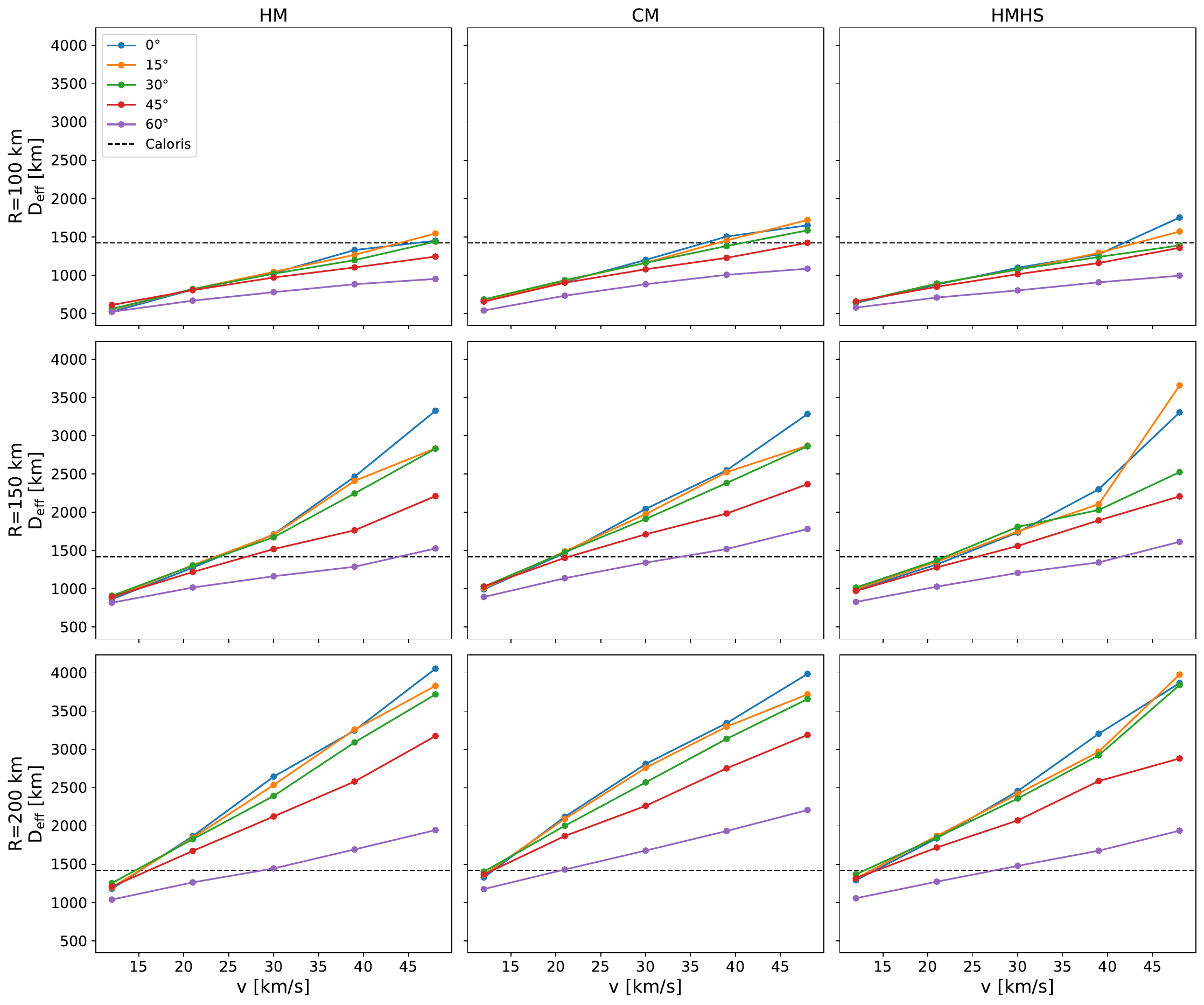}
\caption{Effective basin diameter $D_{\rm eff}$ as a function of impact velocity, for all 225 simulations at 32-million-particle resolution and \SI{104}{\kilo\meter} crust thickness. Rows: impactor radius (\SI{100}{}, \SI{150}{}, \SI{200}{\kilo\meter}). Columns: target thermal profile. Color: impact angle (\SI{0}{\degree} = vertical incidence). The horizontal dashed line marks the observed area-equivalent diameter of the Caloris basin, $D_{\rm eff}\approx\SI{1420}{\kilo\meter}$ \citep{fassettCalorisImpactBasin2009}.}
\label{fig:eff_diameter_overview}
\end{figure*}

\begin{figure*}[ht!]
\centering
\includegraphics[width=0.95\linewidth]{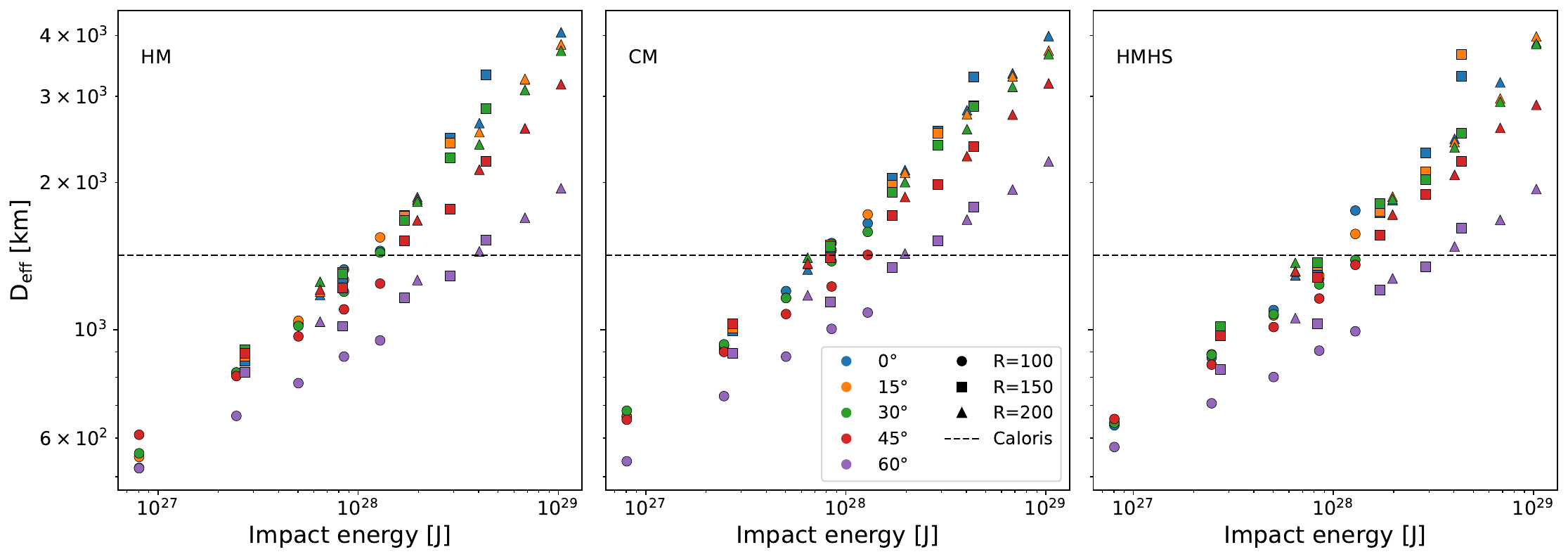}
\caption{Effective basin diameter versus impact energy (log--log), one panel per thermal profile. Color: impact angle. Marker shape: impactor radius. The horizontal dashed line marks the observed area-equivalent diameter of the Caloris basin, $D_{\rm eff}\approx\SI{1420}{\kilo\meter}$ \citep{fassettCalorisImpactBasin2009}.}
\label{fig:crater_size_vs_energy}
\end{figure*}

\begin{figure}[ht!]
\centering
\includegraphics[width=\linewidth]{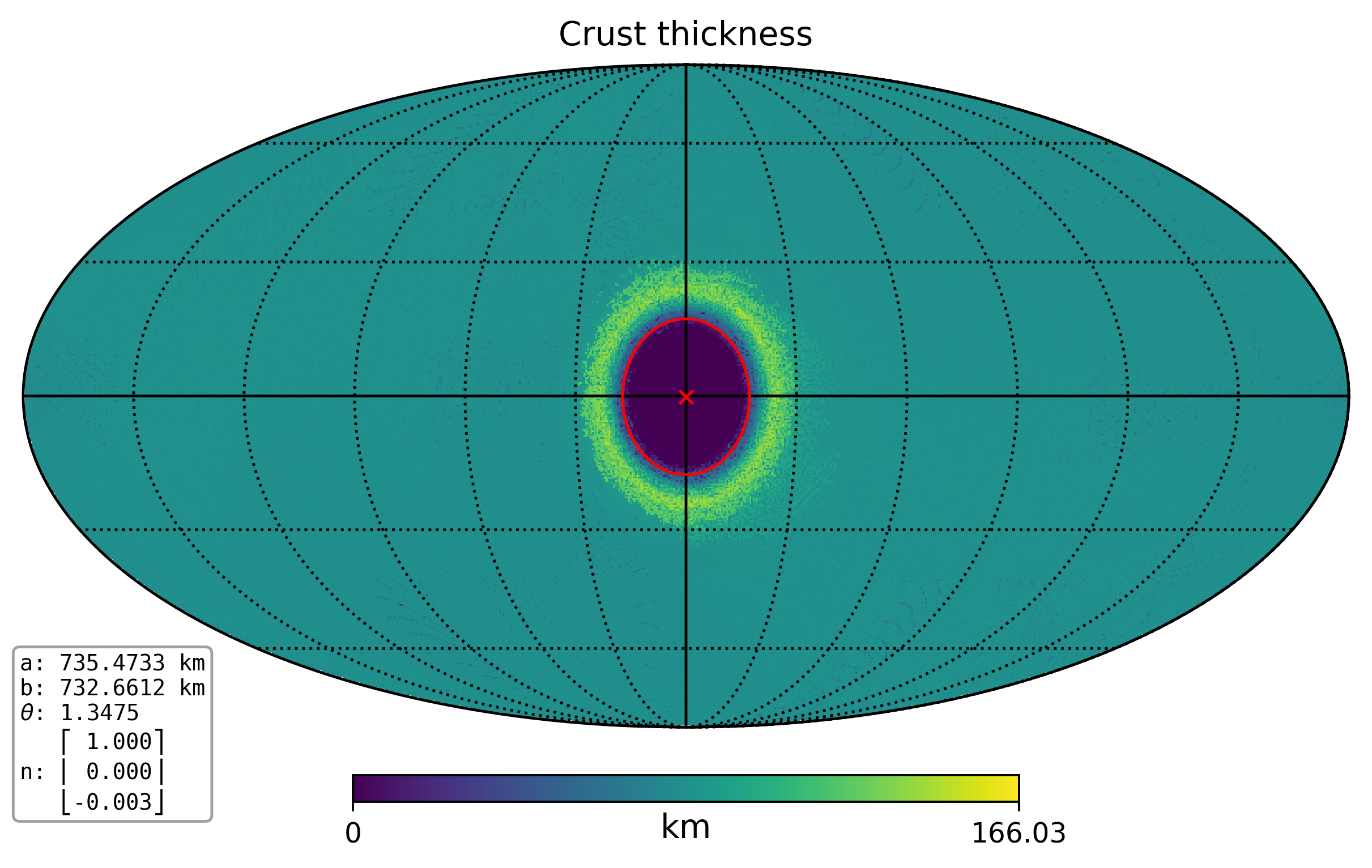}\\
\includegraphics[width=\linewidth]{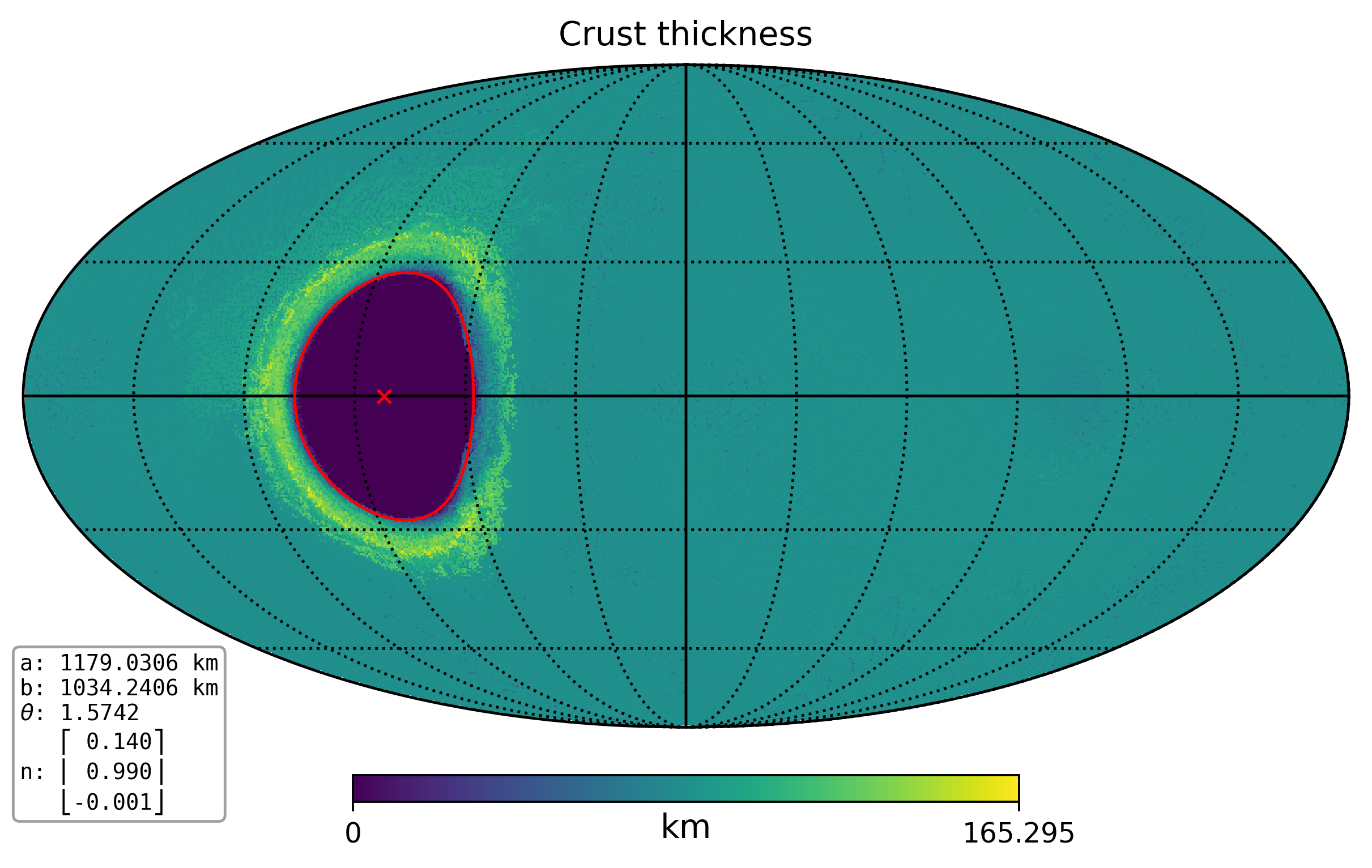}
\caption{Crustal-thickness maps (Mollweide projection) with the deduced basin outline overlaid and the fitted parameters shown in the bottom-left corner, for two 32-million-particle simulations with the CM profile: $R=\SI{150}{\kilo\meter}$, $v=\SI{21}{\kilo\meter\per\second}$, $\theta=\SI{0}{\degree}$ (top) and $R=\SI{200}{\kilo\meter}$, $v=\SI{48}{\kilo\meter\per\second}$, $\theta=\SI{60}{\degree}$ (bottom). The color scale spans from 0 to twice the global median thickness. The construction of the map, the basin classification, and the ellipse fit are described in Appendix~\ref{sec:appendix:Measuring_Crater_Size}.}
\label{fig:crater_map}
\end{figure}

The 225 simulations at 32-million-particle resolution (\SI{104}{\kilo\meter} crust thickness) cover every combination of impactor radius (\SI{100}{}, \SI{150}{}, \SI{200}{\kilo\meter}), impact velocity (\SIrange{12}{48}{\kilo\meter\per\second}), impact angle (\SIrange{0}{60}{\degree}), and target thermal profile (CM, HM, and HMHS; Section~\ref{sec:Models}); the resulting effective diameters are shown in Figure~\ref{fig:eff_diameter_overview}. As our measure of basin size we use the effective basin diameter $D_{\rm eff} = 2\sqrt{ab}$, the diameter of a circle with the same area as the elliptical basin outline with semi-major axis $a$ and semi-minor axis $b$; Appendix~\ref{sec:appendix:Measuring_Crater_Size} describes how $a$ and $b$ are determined from the simulation output. Across the parameter space, $D_{\rm eff}$ ranges from $\approx\SI{550}{\kilo\meter}$ (\SI{100}{\kilo\meter} impactor, \SI{12}{\kilo\meter\per\second}, most oblique case) to $\approx\SI{4000}{\kilo\meter}$ (\SI{200}{\kilo\meter} impactor, \SI{48}{\kilo\meter\per\second}, near-vertical incidence; Figures~\ref{fig:eff_diameter_overview} and \ref{fig:crater_size_vs_energy}). Impact velocity and impactor radius are the two dominant controls on basin size, consistent with their combined role in setting the impact energy (Figure~\ref{fig:crater_size_vs_energy}); impact angle is a secondary but systematic modifier, visible as the color ordering within each panel of Figure~\ref{fig:eff_diameter_overview}. Figure~\ref{fig:crater_map} shows the crustal-thickness maps and fitted basin outlines for two representative cases spanning this range: a moderate basin from a vertical impact and a very large, elongated basin from a fast, highly oblique one.

\subsubsection{Influence of impactor mass, velocity, and impact angle}\label{sec:scaling}
\begin{table*}[ht!]
\centering
\begin{tabular}{llll}
\toprule
Description & Form & Parameters & Key assumption \\
\midrule
Point-source theory & $D \propto R^{2/(2+\mu)}\, v^{2\mu/(2+\mu)}$ & $\mu \in [1/3,\, 2/3]$ & scale-invariant (half-space) target \\
Oblique extension & $\times\, \cos(\theta)^{\gamma}$ & $\gamma \approx \numrange{0.33}{0.44}$ & vertical velocity component controls coupling \\
Energy scaling & $D \propto E^{a}$ & $a$ ($= 1/4$ for $\mu = 2/3$) & point-source limit $\mu = 2/3$ \\
This work & Equation~\eqref{eq:scaling-law} & $D_0$, $\alpha$, $\beta$, $\gamma$ free & none of the above imposed \\
\bottomrule
\end{tabular}
\caption{Scaling descriptions used in this subsection. Point-source theory ties the radius and velocity exponents to a single coupling exponent $\mu$; energy scaling is its $\mu = 2/3$ limit; the oblique-impact extension multiplies either description by a power of $\cos(\theta)$. The empirical fit of this work leaves all exponents free and thus contains each literature description as a testable restriction. Literature values for the exponents are listed in Table~\ref{tab:scaling-fits}.}
\label{tab:scaling-descriptions}
\end{table*}

\begin{table*}[ht!]
\centering
\begin{tabular}{lcccccc}
\toprule
Fit & $\alpha$ (radius) & $\beta$ (velocity) & $\gamma$ (angle) & $D_0$ [\si{\kilo\meter}] & $R^2$ & $n$ \\
\midrule
HM       & \num{1.190\pm0.043} & \num{0.695\pm0.025} & \num{0.569\pm0.047} & \num{1434\pm26} & 0.960 & 75 \\
CM     & \num{1.120\pm0.029} & \num{0.648\pm0.017} & \num{0.539\pm0.032} & \num{1593\pm19} & 0.979 & 75 \\
HMHS   & \num{1.098\pm0.042} & \num{0.628\pm0.024} & \num{0.576\pm0.047} & \num{1505\pm27} & 0.955 & 75 \\
\midrule
Pooled (all profiles) & \num{1.136\pm0.024} & \num{0.657\pm0.014} & \num{0.561\pm0.027} & \num{1509\pm15} & 0.955 & 225 \\
\midrule
Literature ($\mu=0.55$, competent rock) & $\sim\num{0.78}$ & $\sim\num{0.43}$ & 0.33--0.44 & -- & -- & -- \\
\bottomrule
\end{tabular}
\caption{Fitted parameters of Equation~\eqref{eq:scaling-law}, by thermal profile and pooled across all three; $D_0$ is the amplitude of the fitted relation, evaluated at the reference conditions $R=\SI{150}{\kilo\meter}$, $v=\SI{21}{\kilo\meter\per\second}$, $\theta=\SI{0}{\degree}$. The exponents are compared to point-source scaling theory for competent rock ($\mu \approx 0.55$; \citet{holsapplePointSourceSolutions1987,schmidtRecentAdvancesScaling1987,housenEjectaImpactCraters2011}) and the oblique-impact angle exponent commonly used in the literature \citep{collinsEarthImpactEffects2005,elbeshausenScalingObliqueImpacts2009,johnsonSpheruleLayersCrater2016}. Uncertainties are 1$\sigma$ standard errors from the least-squares fit.}
\label{tab:scaling-fits}
\end{table*}

Three scaling descriptions from the literature serve as reference points in this subsection; Table~\ref{tab:scaling-descriptions} summarizes their functional forms, parameters, and key assumptions. Point-source theory predicts power-law dependence of crater size on impactor radius and velocity, with both exponents controlled by a single coupling exponent $\mu$. Its common oblique-impact extension multiplies this by a power of the cosine of the impact angle. Energy scaling collapses the radius and velocity dependence onto the impact energy alone and is the special case of point-source theory with $\mu = 2/3$. Against these references, we fit an empirical relation in which all exponents are free parameters; each literature description is then a testable restriction of this fit. We first present the fit and then compare it to each reference description in turn.

\begin{figure}[ht!]
\centering
\includegraphics[width=\linewidth]{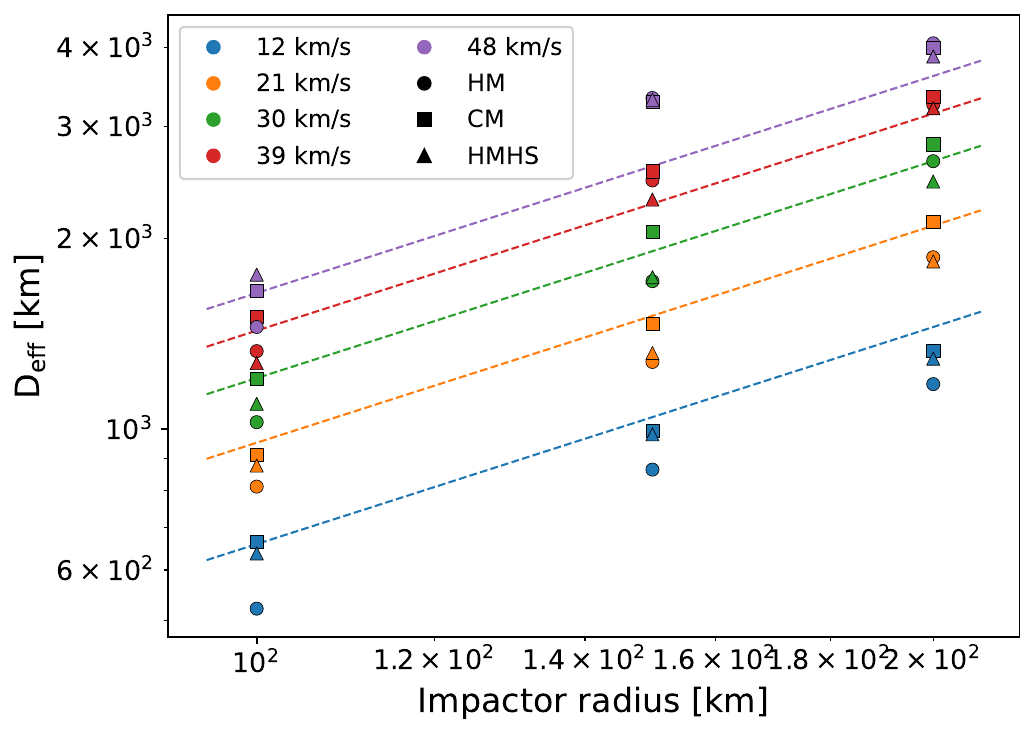}
\caption{Effective basin diameter versus impactor radius at normal incidence (\SI{0}{\degree}), for all 5 impact velocities and all 3 thermal profiles (color: velocity; marker: thermal profile). Dashed lines: pooled power-law fit (Table~\ref{tab:scaling-fits}), evaluated at each velocity.}
\label{fig:crater_size_vs_radius}
\end{figure}

\begin{figure}[ht!]
\centering
\includegraphics[width=\linewidth]{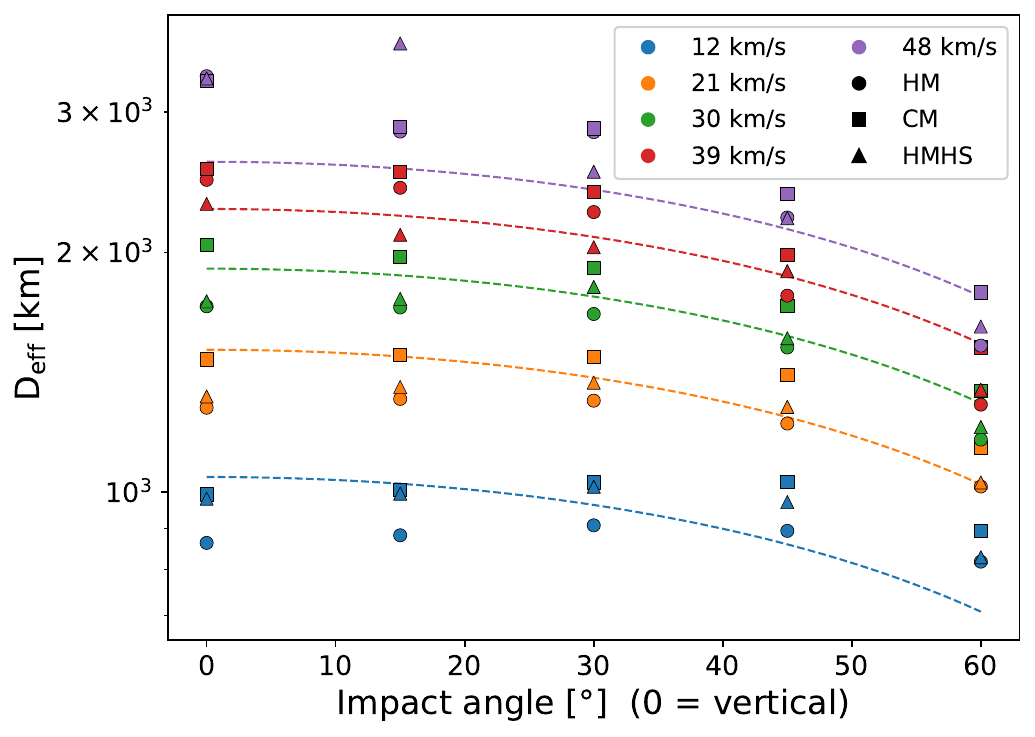}
\caption{Effective basin diameter versus impact angle at $R=\SI{150}{\kilo\meter}$, for all 5 impact velocities and all 3 thermal profiles (color: velocity; marker: thermal profile). Dashed lines: pooled power-law fit (Table~\ref{tab:scaling-fits}), evaluated at each velocity. \SI{0}{\degree} corresponds to vertical incidence.}
\label{fig:crater_size_vs_angle}
\end{figure}

\textbf{Empirical fit --} Impactor mass is not an independent parameter in this study. For a fixed bulk density of $\rho \approx \SI{2.65}{\gram\per\centi\meter\tothe3}$, the three simulated radii correspond to particle-model masses of \SI{1.12e19}{}, \SI{3.78e19}{}, and \SI{8.96e19}{\kilo\gram}; these are approximately one percent above the analytic values for homogeneous spheres because the relaxed bodies are compressed by their own gravity. Because mass and radius are perfectly correlated ($m \propto R^3$) in the design, we report the influence of impactor \emph{radius} throughout rather than attempting to separate a mass effect from a size effect; the two are physically and statistically indistinguishable here (Figure~\ref{fig:crater_size_vs_radius}).

To quantify how basin size depends on impactor radius $R$, impact velocity $v$, and impact angle $\theta$, we fit scaling relations of the form

\begin{equation}
D_{\rm eff} = D_0 \left(\frac{R}{\SI{150}{\kilo\meter}}\right)^{\alpha} \left(\frac{v}{\SI{21}{\kilo\meter\per\second}}\right)^{\beta} \cos(\theta)^{\gamma}\,.\label{eq:scaling-law}
\end{equation}

\noindent This functional form is motivated by point-source crater-scaling theory, conventionally formulated in terms of dimensionless ($\pi$-group) variables, which predicts power-law dependence of crater size on impactor size and velocity \citep{holsapplePointSourceSolutions1987,housenEjectaImpactCraters2011}. It also includes the power-law dependence on the sine of the impact angle commonly used to describe oblique impacts, where the angle is conventionally measured from the surface plane (\SI{90}{\degree}~=~vertical) \citep{collinsEarthImpactEffects2005,elbeshausenScalingObliqueImpacts2009}. Because our simulation grid instead measures the impact angle from the vertical (\SI{0}{\degree}~=~normal incidence), we write the angular dependence as $\cos(\theta)^{\gamma}$, so that the fitted exponent $\gamma$ remains directly comparable to published values. The fits are performed by ordinary least squares in log--log space, separately for each thermal profile and pooled across all three (Table~\ref{tab:scaling-fits}; fit lines in Figures~\ref{fig:crater_size_vs_radius}~and~\ref{fig:crater_size_vs_angle}).

The pooled fit ($n=225$, $R^2 = 0.955$) gives

\begin{align}
D_{\rm eff} &= \SI{1509\pm15}{\kilo\meter} \left(\frac{R}{\SI{150}{\kilo\meter}}\right)^{1.14\pm0.02} \nonumber\\
&\quad\times\left(\frac{v}{\SI{21}{\kilo\meter\per\second}}\right)^{0.66\pm0.01} \cos(\theta)^{0.56\pm0.03}\,.
\end{align}

\noindent The three thermal profiles individually return very similar exponents (radius: 1.10--1.19; velocity: 0.63--0.70; angle: 0.54--0.58) but clearly distinct amplitudes: $D_0$ spans \SIrange{1434}{1593}{\kilo\meter}, an \SI{11}{\percent} spread well beyond the fit uncertainties and largest for the cold-mantle profile (Table~\ref{tab:scaling-fits}). The target's thermal structure therefore mainly rescales the overall basin size rather than changing its functional dependence on the impactor properties. The direction of this rescaling is notable: the cold-mantle profile, the strongest target under thermal softening, produces the largest basins. The final basin size measured from the crust is thus not controlled by excavation efficiency alone; a plausible interpretation is that hotter, weaker targets accommodate more post-impact inflow and collapse of the cavity, reducing the final crustal basin, although we have not isolated the mechanism here.

We note that $D_0$ exceeds the measured diameter at the reference grid point by \SIrange{8}{14}{\percent} for all profiles. This is because the measured $D_{\rm eff}$ peaks at moderate obliquity rather than at vertical incidence, so the monotonic $\cos(\theta)^{\gamma}$ form places the fitted surface above the measured vertical-incidence values. $D_0$ should therefore be read as the amplitude of the fitted relation, not as the measured diameter at the reference conditions.

\textbf{Comparison with point-source and oblique-impact scaling --} For comparison, point-source scaling for competent, non-porous rock with a velocity-scaling exponent $\mu \approx 0.55$ predicts $D \propto R^{0.78}\,v^{0.43}$ at fixed angle \citep{holsapplePointSourceSolutions1987,housenEjectaImpactCraters2011}, while the oblique-impact scaling commonly used in the literature corresponds to $D(\theta)/D(\SI{0}{\degree}) = \cos(\theta)^{0.33\text{--}0.44}$ in our angle convention \citep{collinsEarthImpactEffects2005,elbeshausenScalingObliqueImpacts2009,johnsonSpheruleLayersCrater2016}. Our fitted exponents exceed these literature values on all three variables: basin size in our simulations responds more strongly to impactor radius, velocity, and impact angle than idealized simple-crater scaling predicts. This is plausible for two reasons. First, the literature exponents are calibrated primarily on small, simple (bowl-shaped) craters and laboratory experiments, whereas the Caloris-forming impact produces a large, complex, multi-ring basin in which collapse and megaregolith/thermal-weakening effects can amplify the sensitivity of the final basin to the impact conditions. Second, Mercury's differentiated, radially varying thermal structure is not represented in the generic point-source formulation. We regard the qualitative agreement (same sign, same order of magnitude, and comparable relative ordering of the three exponents) as evidence that the simulations behave physically sensibly, while the quantitative offset highlights the difference between idealized simple-crater scaling and basin-scale, thermally structured cratering.

\textbf{The coupling exponent $\mu$ --} The fitted exponents can be related to the velocity-scaling exponent $\mu$ of point-source theory, and this comparison localizes where the point-source description fails. In the gravity regime, the point-source assumption fixes both exponents in terms of $\mu$ alone, $D \propto R^{2/(2+\mu)}\, v^{2\mu/(2+\mu)}$, so the radius exponent $\alpha$ and velocity exponent $\beta$ are not independent: they must satisfy $\alpha + \beta/2 = 1$, and their ratio gives the coupling exponent, $\mu = \beta/\alpha$ \citep{holsapplePointSourceSolutions1987,housenEjectaImpactCraters2011}. Our pooled fit gives $\alpha + \beta/2 = \num{1.47\pm0.02}$; the point-source constraint is clearly violated, and no single value of $\mu$ can reproduce both exponents. The ratio, however, remains meaningful. Rewriting the fitted relation as $D_{\rm eff} \propto \left(R\, v^{0.58}\right)^{1.14}$ shows that the basin size still responds to a coupling-parameter combination $C = R\, v^{\mu}$, with $\mu = \beta/\alpha = \num{0.58\pm0.01}$. This value lies within the dimensionally allowed range between momentum coupling ($\mu = 1/3$) and energy coupling ($\mu = 2/3$) and is close to the $\mu \approx 0.55$ measured for competent rock and wet soils \citep{schmidtRecentAdvancesScaling1987,housenEjectaImpactCraters2011}. What deviates from point-source theory is instead the power to which this combination enters: we find $\num{1.14\pm0.02}$ where the point-source prediction is $2/(2+\mu) \approx 0.78$. The violation is thus localized in the size dependence. This is the expected failure mode when the point-source premise of a scale-invariant target does not hold: our impactors vary by a factor of two in radius against a crust thickness, mantle depth, and planetary curvature of fixed scale, so impactors of different sizes do not see self-similar targets. The velocity dependence at fixed radius is immune to this effect, because impactors of every velocity encounter the same target structure, which is why the coupling exponent extracted from the ratio remains physical while the overall power absorbs the deviation.

\textbf{Energy scaling --} A complementary way to present the same information, common for planetary-scale basins, is to scale the basin size against the impact energy alone. This is the approach taken by \citet{marinovaGeophysicalConsequencesPlanetaryscale2011}, whose SPH survey of planetary-scale impacts into a Mars-like planet, performed without material strength, spans \SIrange{2e27}{6e29}{\joule}, overlapping the $\approx\SIrange{8e26}{1e29}{\joule}$ of our survey. For vertical incidence, our data justify this approach: a single-variable fit gives $D_{\rm eff} \propto E^{0.40\pm0.01}$ (pooled across the three thermal profiles) and describes the vertical-incidence data as well as the unrestricted two-parameter fit in radius and velocity ($R^2 = 0.974$ in both cases; Figure~\ref{fig:crater_size_vs_energy}). The reason is a mild angle dependence of the fitted exponents: restricted to vertical incidence, the fit returns $\alpha = \num{1.22\pm0.05}$ and $\beta = \num{0.80\pm0.03}$, whose ratio $\mu = \num{0.66\pm0.03}$ is consistent with pure energy coupling ($\mu = 2/3$), so at vertical incidence, impact energy is a nearly sufficient variable. The pooled $\mu = \num{0.58\pm0.01}$ of the full survey is lower because the effective coupling softens with increasing obliquity. The energy exponent itself is again markedly steeper than the point-source expectation, which predicts $D \propto E^{1/4}$ for $\mu = 2/3$ in the gravity regime. Once all angles are pooled, energy and a $\cos(\theta)$ factor alone are no longer sufficient: rewriting Equation~\eqref{eq:scaling-law} in terms of energy and velocity gives $D_{\rm eff} \propto E^{0.38}\, v^{-0.10}$, so at fixed impact energy, larger and slower impactors produce slightly larger basins, and the corresponding restriction $\mu = 2/3$, imposed on the full data set via $D_{\rm eff} \propto E^{a}\cos(\theta)^{c}$, is rejected at $F = 21.1$ ($p \approx 7\times10^{-6}$). The systematic deviation of both descriptions from half-space, point-source expectations is consistent with the conclusions of \citet{marinovaGeophysicalConsequencesPlanetaryscale2011}, who attributed such deviations to surface curvature, radial gravity, the large impactor-to-planet size ratio, and the deeper penetration of the impactor, the same target-scale effects that we identify above as breaking the point-source premise. The $(R, v, \theta)$ decomposition of Equation~\eqref{eq:scaling-law} nevertheless remains the more diagnostic representation, and it is the one used for the reference fit lines in Figures~\ref{fig:crater_size_vs_radius}~and~\ref{fig:crater_size_vs_angle}.

\textbf{Non-monotonic angle dependence --} We also note that the angle dependence of $D_{\rm eff}$ is not strictly monotonic: in 25 of the 45 (profile, radius, velocity) combinations, the effective diameter peaks at an intermediate angle (typically \SIrange{15}{30}{\degree} from vertical) rather than decreasing monotonically from vertical to oblique incidence. We interpret this as reflecting basin elongation: with increasing obliquity, the semi-major axis initially grows even as the total excavated volume falls, before energy-coupling losses dominate at higher obliquity. We do not attribute the effect to numerical noise, since it is systematic across most of the parameter grid. It is visible as a modest deviation of the scatter from the fitted $\cos(\theta)^{0.56}$ curve in Figure~\ref{fig:crater_size_vs_angle}, particularly at low to intermediate angles.

\subsubsection{Sensitivity to numerical resolution and crustal thickness}\label{sec:resolution_sensitivity}
\begin{table}[ht!]
\centering
\begin{tabular}{lccc}
\toprule
Thermal profile & Resolution & Crust thickness & $\Delta D_{\rm eff}$ \\
\midrule
HM     & 32M   & \SI{104}{\kilo\meter} & baseline \\
                 & 256M  & \SI{72.5}{\kilo\meter} & \num[explicit-sign = +]{+2.74}\% \\
                 & 256M  & \SI{104}{\kilo\meter}  & \num{-4.12}\% \\
\midrule
CM   & 32M   & \SI{104}{\kilo\meter}  & baseline \\
                 & 256M  & \SI{72.5}{\kilo\meter} & \num[explicit-sign = +]{+6.78}\% \\
                 & 256M  & \SI{104}{\kilo\meter}  & \num[explicit-sign = +]{+17.19}\% \\
                 & 2048M & \SI{40}{\kilo\meter}   & \num[explicit-sign = +]{+23.11}\% \\
\midrule
HMHS & 32M   & \SI{104}{\kilo\meter} & baseline \\
                 & 256M  & \SI{72.5}{\kilo\meter} & \num[explicit-sign = +]{+0.75}\% \\
                 & 256M  & \SI{104}{\kilo\meter}  & \num{-9.84}\% \\
\bottomrule
\end{tabular}
\caption{Resolution and crust-thickness sensitivity at $v=\SI{21}{\kilo\meter\per\second}$, $\theta=\SI{0}{\degree}$ (vertical), $R=\SI{150}{\kilo\meter}$. Percentages are relative to the 32-million-particle, \SI{104}{\kilo\meter}-crust baseline for each thermal profile.}
\label{tab:resolution_crust_sensitivity}
\end{table}

\begin{table*}[ht!]
\centering
\begin{tabular}{lccc}
\toprule
Decoupled effect & HM & CM & HMHS \\
\midrule
Crust $\SI{104}{\kilo\meter}\,\rightarrow\,\SI{72.5}{\kilo\meter}$ at 256M & \num[explicit-sign = +]{+7.15}\% & \num{-8.88}\% & \num[explicit-sign = +]{+11.75}\% \\
Resolution 32M$\,\rightarrow\,$256M at \SI{104}{\kilo\meter} & \num{-4.12}\% & \num[explicit-sign = +]{+17.19}\% & \num{-9.84}\% \\
\bottomrule
\end{tabular}
\caption{Single-variable sensitivities of $D_{\rm eff}$, obtained from the runs of Table~\ref{tab:resolution_crust_sensitivity} by comparing pairs that differ in only crust thickness or only resolution. In both decompositions, the CM profile responds with the opposite sign to the other two profiles.}
\label{tab:decoupled_sensitivity}
\end{table*}

\begin{figure}[t]
\centering
\includegraphics[width=\linewidth]{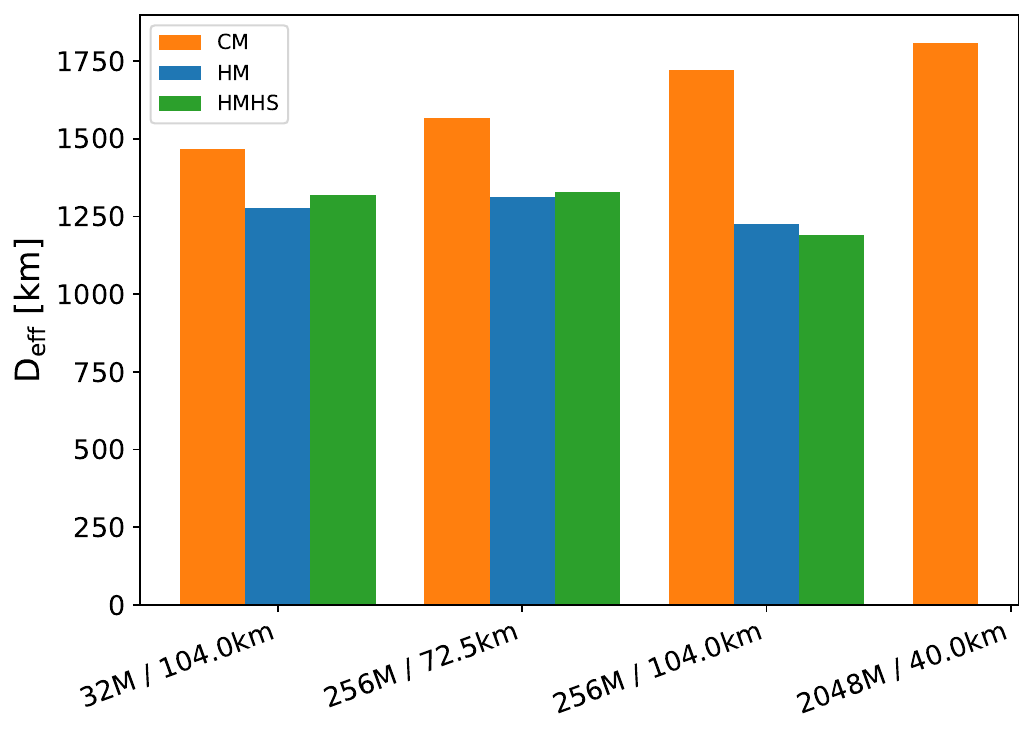}
\caption{Effective basin diameter for the resolution- and crust-thickness sensitivity runs of Table~\ref{tab:resolution_crust_sensitivity}, at $v=\SI{21}{\kilo\meter\per\second}$, $\theta=\SI{0}{\degree}$, $R=\SI{150}{\kilo\meter}$. Bars grouped by (resolution / crust thickness) combination; color: thermal profile.}
\label{fig:resolution_crust_sensitivity}
\end{figure}

To assess numerical convergence and the influence of crustal thickness independently of the impact parameters, we ran a limited set of additional simulations at higher particle resolution (256 million and 2 billion particles) and reduced crustal thickness (\SI{72.5}{\kilo\meter} and \SI{40}{\kilo\meter}, versus the standard \SI{104}{\kilo\meter}), all at fixed impact conditions ($v=\SI{21}{\kilo\meter\per\second}$, $\theta=\SI{0}{\degree}$ [vertical], $R=\SI{150}{\kilo\meter}$; Table~\ref{tab:resolution_crust_sensitivity}, Figure~\ref{fig:resolution_crust_sensitivity}). Table~\ref{tab:resolution_crust_sensitivity} lists the change in $D_{\rm eff}$ relative to the 32-million-particle, \SI{104}{\kilo\meter}-crust baseline of each thermal profile. The 256M/\SI{72.5}{\kilo\meter} runs differ from this baseline in both resolution \emph{and} crust thickness, so the two effects must be separated by comparing runs that differ in only one variable. The effect of crustal thinning at fixed resolution follows from comparing the \SI{72.5}{\kilo\meter}- and \SI{104}{\kilo\meter}-crust runs at 256M particles. The effect of resolution alone follows from comparing the 256M and 32M runs at fixed \SI{104}{\kilo\meter} crust thickness (Table~\ref{tab:decoupled_sensitivity}).

Neither effect is consistent in sign across the thermal profiles. Thinning the crust from \SI{104}{} to \SI{72.5}{\kilo\meter} at fixed 256M resolution increases $D_{\rm eff}$ by 7\% and 12\% for the HM and HMHS profiles, respectively, but \emph{decreases} it by 9\% for the CM profile. Conversely, increasing the resolution from 32M to 256M particles at fixed \SI{104}{\kilo\meter} crust thickness decreases $D_{\rm eff}$ by 4\% and 10\% for HM and HMHS, but \emph{increases} it by 17\% for CM. In both decompositions, the CM profile responds with the opposite sign to the other two. This indicates that the basin size has not reached a resolution-independent regime, at least for this impact condition and particle-count range, and that both sensitivities depend on the target's thermal structure. The single simulation at the highest resolution and the thinnest, observationally motivated crust (2 billion particles, \SI{40}{\kilo\meter} crust, CM only) yields a $D_{\rm eff}$ 23\% above the corresponding baseline. Since no 2-billion-particle run with a thicker crust (nor a 256M run with \SI{40}{\kilo\meter} crust) is available, the resolution and crust-thickness contributions to this change cannot be separated. Taken together, the individual effects are of order \SIrange{4}{17}{\percent} in $D_{\rm eff}$, which we adopt as the characteristic systematic uncertainty associated with resolution and crust thickness in this study. We note in addition that \SI{40}{\kilo\meter} lies at the upper end of current observational estimates of Mercury's mean crustal thickness (\SIrange{26}{50}{\kilo\meter}; Section~\ref{sec:Models}). Given the measured sensitivity of $D_{\rm eff}$ to crustal thickness, a still thinner true crust would plausibly change the basin size further, in a direction and by an amount our present series cannot constrain. Because only one run is available per condition, we treat these percentages as indicative rather than as converged trends, and recommend dedicated single-variable series (varying particle count at fixed crust thickness, and vice versa, for all three thermal profiles) before either sensitivity is treated as fully characterized.

\subsubsection{Implications for Caloris basin formation}\label{sec:caloris_implications}
The size of the Caloris basin is commonly quoted as a rim-to-rim diameter of approximately \SI{1550}{\kilo\meter} \citep{murchieGeologyCalorisBasin2008}. The basin rim is, however, slightly elliptical: a best-fit ellipse gives major and minor axes of \SI{1525}{} and \SI{1315}{\kilo\meter} \citep{fassettCalorisImpactBasin2009}, corresponding to an area-equivalent diameter of $D_{\rm eff}\approx\SI{1420}{\kilo\meter}$, the quantity directly comparable to the effective diameters measured from our simulations. Additional candidate basin rings have been reported out to $\sim\SI{3700}{\kilo\meter}$ \citep{spudisStratigraphyGeologicHistory1988}. Within our simulation grid, the observed basin size is matched by a broad swath of (radius, velocity, angle) combinations rather than by a single, narrowly constrained impactor. For example, impactors of \SI{150}{\kilo\meter} radius at velocities of \SIrange{21}{30}{\kilo\meter\per\second} produce effective diameters ranging from $\approx\SI{1020}{\kilo\meter}$ (most oblique case) to $\approx\SI{2040}{\kilo\meter}$ (vertical incidence; Figures~\ref{fig:eff_diameter_overview} and \ref{fig:crater_size_vs_radius}). Within this family, the observed $D_{\rm eff}\approx\SI{1420}{\kilo\meter}$ is matched only at oblique incidence: at \SI{21}{\kilo\meter\per\second} by the cold-mantle profile at $\theta\approx\SI{42}{\degree}$, and at \SI{30}{\kilo\meter\per\second} by all three thermal profiles at $\theta\approx\SIrange{49}{57}{\degree}$. Both hot-mantle profiles fall short of the observed diameter at this impactor size and \SI{21}{\kilo\meter\per\second} for all impact angles ($D_{\rm eff}\lesssim\SI{1375}{\kilo\meter}$). Given the \SIrange{4}{17}{\percent} systematic uncertainty established in Section~\ref{sec:resolution_sensitivity}, these matching angles are indicative rather than sharp; the near-vertical cold-mantle cases at \SI{21}{\kilo\meter\per\second}, for instance, lie within \SI{5}{\percent} of the observed value. The preference for oblique incidence within this family is nonetheless notable, as an oblique Caloris-forming impact is independently suggested by the ellipticity of the observed basin rim \citep{fassettCalorisImpactBasin2009}. Both aspects of this family are dynamically plausible a priori. For an isotropic impactor flux, the probability distribution of impact angles follows $\sin(2\theta)$ \citep[a classical result reviewed in][]{pierazzoUnderstandingObliqueImpacts2000}, which peaks at \SI{45}{\degree} (measured from vertical, the convention used here). The moderately-to-strongly oblique geometries required to match the observed diameter are therefore more probable than near-vertical incidence. Likewise, velocities of \SIrange{21}{30}{\kilo\meter\per\second} lie below the mean impact velocity at Mercury (\SIrange{30}{43}{\kilo\meter\per\second}; Section~\ref{sec:Initial_conditions}), but the underlying velocity distribution is broad \citep{lefeuvreNonuniformCrateringMoon2011} and a substantial fraction of Mercurian impacts occurs in this range. Neither the velocity nor the angle of the inferred family therefore requires an atypical impactor trajectory at this impactor size; for other sizes, the family shifts along the fitted relation (Equation~\eqref{eq:scaling-law}), with larger impactors matching at slower velocities or more oblique incidence and smaller impactors requiring faster or more nearly vertical impacts. The slower velocities implied by the crust-thickness correction of Section~\ref{sec:resolution_sensitivity} remain within the range sampled by the underlying distribution, though further into its lower tail.

This mapping is, however, defined at the survey conditions of \SI{104}{\kilo\meter} crust thickness and 32-million-particle resolution, and the sensitivity analysis of Section~\ref{sec:resolution_sensitivity} shows that it inherits the corresponding systematic uncertainty. In particular, the single simulation at the observationally motivated crust thickness of \SI{40}{\kilo\meter} (2 billion particles, CM profile) yields a $D_{\rm eff}$ \SI{23}{\percent} larger than its survey-condition counterpart at identical impact parameters. If a correction of this magnitude applies across the grid, the observed $D_{\rm eff}\approx\SI{1420}{\kilo\meter}$ corresponds to a survey-condition diameter of only $\approx\SI{1150}{\kilo\meter}$, shifting the entire admissible region toward smaller or slower impactors. Propagated through the pooled scaling relation (Equation~\eqref{eq:scaling-law}), a reduction of the required diameter by a factor of 1.23 corresponds to a factor $(1.23)^{-1/1.14} \approx 0.83$ in impactor radius or a factor $(1.23)^{-1/0.66} \approx 0.73$ in impact velocity at any fixed point of the admissible region (e.g., $\SI{150}{\kilo\meter} \rightarrow \approx\SI{125}{\kilo\meter}$ at fixed velocity and angle). Because this correction is anchored by a single high-resolution run of one thermal profile (Section~\ref{sec:resolution_sensitivity}), we regard the shift as a systematic uncertainty band on the inferred region rather than a revised best estimate.

Our impactors have radii of \SIrange{100}{200}{\kilo\meter} (diameters of \SIrange{200}{400}{\kilo\meter}; the lower bound is set by the impactor-resolution requirement of Section~\ref{sec:Models}). They are larger than the \SIrange{80}{150}{\kilo\meter}-diameter bodies of previous hydrocode studies \citep{ivanovBasinformingImpactsReconnaissance2010,potterBasinFormationMercury2017,gosselinCrustalBlockMuted2023} and the \SIrange{100}{200}{\kilo\meter}-diameter projectiles of interior-response models \citep{robertsEffectCalorisImpact2012}, but smaller than the $\sim\SI{525}{\kilo\meter}$-diameter impactor suggested by an analysis of the basin-related sculpture \citep{schultzSizesNatureBasin2017}. In particular, the \SI{300}{\kilo\meter}-diameter impactors that reproduce the observed basin size in our simulations are a factor of \numrange{2}{2.5} larger than the best-fit impactors of axisymmetric hydrocode models at comparable velocities \citep{potterBasinFormationMercury2017,gosselinCrustalBlockMuted2023}. Whether this difference reflects the different codes, the different material and strength models and their parameters, the inclusion of oblique geometries, or the different basin-size metrics (crustal-structure fitting versus the crust-particle-based $D_{\rm eff}$ used here) remains to be established.

The role of the target's thermal state deserves emphasis in this comparison. \citet{gosselinCrustalBlockMuted2023} find that reproducing Caloris' observed crustal structure in axisymmetric models requires pre-impact near-surface thermal gradients of \SIrange{22}{30}{\kelvin\per\kilo\meter}, hotter than previously inferred, with the thermal state controlling whether the basin forms with multiring or megabasin morphology. Our simulations probe the thermal dependence through a different observable and find it to be similarly consequential. At fixed impact conditions, the effective basin diameter differs between our thermal profiles by \SI{11}{\percent} in the fitted amplitudes $D_0$ (Table~\ref{tab:scaling-fits}) and by up to \SI{15}{\percent} in the diameters measured at the reference conditions. Moreover, the sensitivity of the basin size to numerical resolution and crustal thickness itself changes sign with the thermal profile (Section~\ref{sec:resolution_sensitivity}). At face value, the two constraints pull in different directions: our cold-mantle profile produces the largest basins at fixed impact conditions, and therefore requires the least energetic impactor to match the observed Caloris diameter, whereas the crustal-structure fitting of \citet{gosselinCrustalBlockMuted2023} favors hot targets. However, the two studies constrain different aspects of the event (final basin diameter versus post-collapse crustal architecture) with different thermal parameterizations (layered isentropes with boundary jumps versus conductive gradients). A direct reconciliation therefore requires measuring the crustal structure, not only the diameter, in three-dimensional simulations of the type presented here. Together, the two studies nevertheless agree on the central point: the pre-impact thermal state of Mercury is a first-order control on the outcome of the Caloris-forming impact, comparable in importance to the impact parameters themselves.

The basin diameter alone therefore cannot uniquely constrain the Caloris-forming impactor; the scaling relations derived here instead provide a physically grounded mapping between the observed basin size and the admissible region of impactor parameter space, subject to the resolution- and crust-thickness systematics quantified in Section~\ref{sec:resolution_sensitivity}.

\subsection{The Caloris Gravity Anomaly}\label{sec:gravity}

Radio tracking of the MESSENGER spacecraft revealed that Mercury's northern hemisphere hosts several large gravity anomalies with amplitudes of order \SI{100}{\milli\Gal} \citep{smithGravityFieldInternal2012}. Among these, the Caloris basin stands out as the only unambiguous positive free-air gravity anomaly associated with an impact basin. \citet{smithGravityFieldInternal2012} identify Caloris as the only definitive mascon basin in their harmonic solution, and \citet{jamesSupportLongwavelengthTopography2015} find that, among Mercury's regions of high-admittance topography, only Caloris features a mantle anomaly centered on an identifiable impact basin. A positive anomaly near the Budh and Sobkou basins may constitute an additional basin-associated mascon \citep{smithGravityFieldInternal2012,mazaricoGravityFieldOrientation2014}, but it is offset from the basin centers and emerges clearly only after correction for a regional topographic rise, leaving its association with either basin ambiguous.

Such an anomaly had in fact been anticipated on geological grounds before the MESSENGER gravity measurement. Based on the pattern of thrust and normal faulting mapped within and around the basin, \citet{kennedyMechanismsFaultingCaloris2008} predicted a central positive free-air anomaly (a mascon) flanked by an annular gravity high, with a ring of comparatively lower gravity in between, provided the Caloris interior had reached isostatic equilibrium before its partial volcanic infilling. This prediction was subsequently borne out, at least for the central positive anomaly, by the observed field \citep{smithGravityFieldInternal2012}.

All simulations analyzed in this and the following section are vertical ($\theta=\SI{0}{\degree}$) impacts, whereas the impact-angle constraint derived in Section~\ref{sec:caloris_implications} and the ellipticity of the observed basin rim \citep{fassettCalorisImpactBasin2009} both point to oblique incidence for the Caloris-forming event. For an oblique impact, the region of maximum mantle thinning and any associated core deformation would be displaced downrange of the geometric basin center and azimuthally asymmetric, rather than forming the axisymmetric dome found here. The qualitative conclusion, that the impact produces a superisostatic excess of dense material beneath the basin, should be robust to moderate obliquity; the geometry, amplitude, and position of the deformation should, however, be regarded as an idealized, axisymmetric limiting case.

\subsubsection{Hypotheses for the origin of the anomaly}\label{sec:gravity_hypotheses}
The classical explanation for lunar and Mercurian mascons invokes \emph{superisostatic uplift of dense mantle material}: excavation of the basin removes low-density crust and places the thinned crust directly against the uplifted mantle. If this mantle material cannot subside back to an isostatically compensated state, for example because a sufficiently rigid lithosphere forms before full relaxation is complete, the resulting excess of dense material near the surface produces a positive gravity anomaly \citep{meloshOriginLunarMascon2013}. \citet{meloshOriginLunarMascon2013} showed that this mechanism, driven by the impact-heated melt pool cooling and contracting beneath an already-solidified lithosphere, can by itself generate mascons without requiring any later volcanic loading. More recently, \citet{gosselinFormationCalorisBasins2025} applied a similar post-impact thermal-contraction model specifically to Caloris and found that contraction of the impact-heated melt pool beneath a newly formed lithosphere draws the surrounding mantle material upward, producing a free-air gravity signature consistent with the observed anomaly without invoking the rigid, strengthless melt pool of earlier work.

A second, historically important class of hypotheses instead attributes positive basin-centered gravity anomalies to a \emph{buried, anomalously dense body} near the impact site. This idea traces back to early interpretations of the lunar mascons as buried remnants of dense impactor material \citep{mullerMasconsLunarMass1968,ureyMasconsHistoryMoon1968,wiseMasconsStructuralRelief1970,meloshOriginLunarMascon2013,dombardOriginMasconBasins2013}. It has since been superseded, for both the Moon and Mercury, by the mantle-uplift/thermal-contraction picture above, largely because forward gravity models based on mantle doming and lithospheric flexure alone reproduce the observed anomalies without requiring a compositionally distinct dense body \citep{meloshOriginLunarMascon2013,dombardOriginMasconBasins2013}. Nonetheless, Mercury's impactors strike at unusually high velocity and could plausibly retain a denser, metal-rich component. Buried impactor material therefore remains a hypothesis worth testing directly against impact-simulation output, rather than dismissing by analogy with the Moon alone.

\subsubsection{Buried impactor material} \label{sec:gravity_impactor}
We first tested whether a buried remnant of impactor material could plausibly contribute to the positive anomaly. Impactor material is unambiguously identified in our simulations by mapping each particle back to the initial conditions via its unique particle ID, so the fate of the impactor can be measured directly. We find that no condensed impactor material remains in the basin region: all impactor material located within the basin at $t=\SI{1.77}{\hour}$, the end of the 2-billion-particle simulation, is vaporized. A vapor plume cannot constitute a buried remnant: it expands, escapes, or recondenses as dispersed fallback (including the antipodal convergence quantified in Section~\ref{sec:antipode_mass}). The premise of the buried-impactor hypothesis, a coherent dense body emplaced at depth, therefore fails at the first step in our simulations. Taken at face value, this argues against the buried-impactor hypothesis for the anomaly, consistent with the mantle-uplift/thermal-contraction picture favored in the literature \citep{meloshOriginLunarMascon2013,gosselinFormationCalorisBasins2025}. This conclusion applies to the velocity range surveyed here ($v_{\rm imp}\geq\SI{12}{\kilo\meter\per\second}$). At impact velocities approaching Mercury's escape velocity of $\approx\SI{4.3}{\kilo\meter\per\second}$, shock vaporization would be far less complete and a coherent buried remnant might survive; such slow impacts are, however, rare for the heliocentric impactor population at Mercury (Section~\ref{sec:Initial_conditions}).

This result comes with two caveats. First, the classification of particles as vaporized is sensitive to the equation of state and to the interface and free-surface correction scheme used in the code, as discussed for the antipodally converging material in Section~\ref{sec:antipode_mass}. This caveat is, however, largely inconsequential here. The impactor in our simulations is undifferentiated and shares the equation of state, density, and strength model of the target crust. Even if some impactor material in the basin region were in fact condensed, it would therefore be crust-like material with no density contrast, and hence no gravity signal, against its surroundings. This same property is the second, structural caveat: with no compositional distinction between impactor and target, our simulations can rule out a buried remnant of \emph{crust-like} impactor material, but cannot address the gravitational signature of a compositionally distinct one. A real Mercurian impactor may plausibly have been differentiated, with a denser metallic core of its own; iron is also more resistant to shock vaporization than the rocky material simulated here, so the wholesale vaporization found for our quartz impactors need not apply to a metallic core component. Whether such a dense core fragment could survive burial beneath the basin, and what anomaly it would produce, therefore remains an open question for simulations with differentiated impactors (Section~\ref{sec:Summary_and_Conclusions}) rather than one our results rule out.

\subsubsection{Mantle thinning and core deformation}\label{sec:gravity_mantle}
\begin{figure*}[ht!]
\centering
\includegraphics[width=0.95\linewidth]{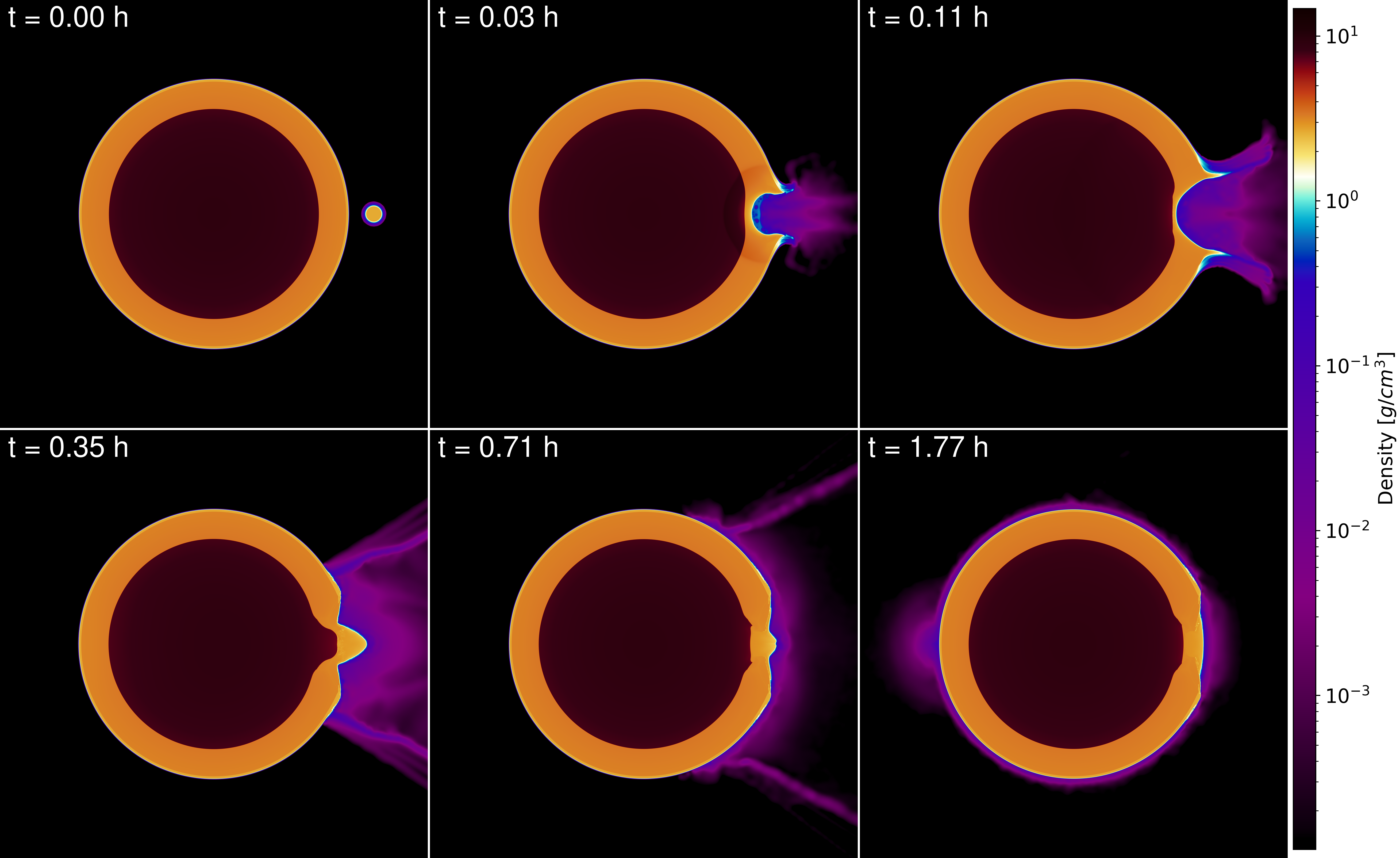}
\caption{Time series of meridional density slices through the impact site for the $R_{\rm imp}=\SI{150}{\kilo\meter}$, $v_{\rm imp}=\SI{21}{\kilo\meter\per\second}$, $\theta=\SI{0}{\degree}$ simulation with the CM profile at 2B particles, illustrating the progressive thinning of the mantle column beneath the basin and the development of a local upward bulge (dome) of the iron core beneath the impact site. An animation of this impact can be found at \href{https://youtu.be/xxzwNLQbAHE}{youtu.be/xxzwNLQbAHE}.}
\label{fig:Impact_Composite_Figure}
\end{figure*}

\begin{figure}[ht!]
\centering
\includegraphics[width=\linewidth]{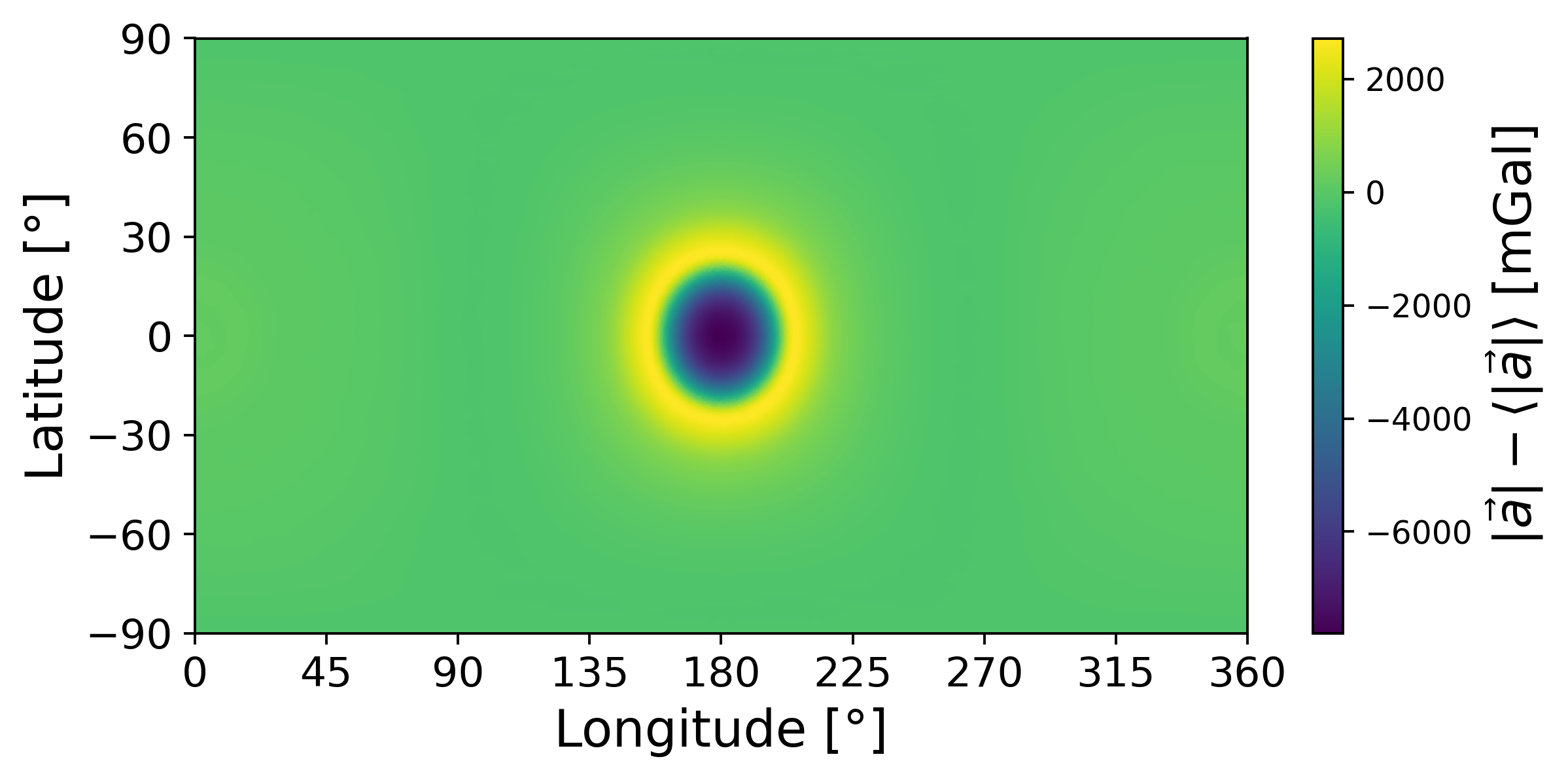}\\
\includegraphics[width=\linewidth]{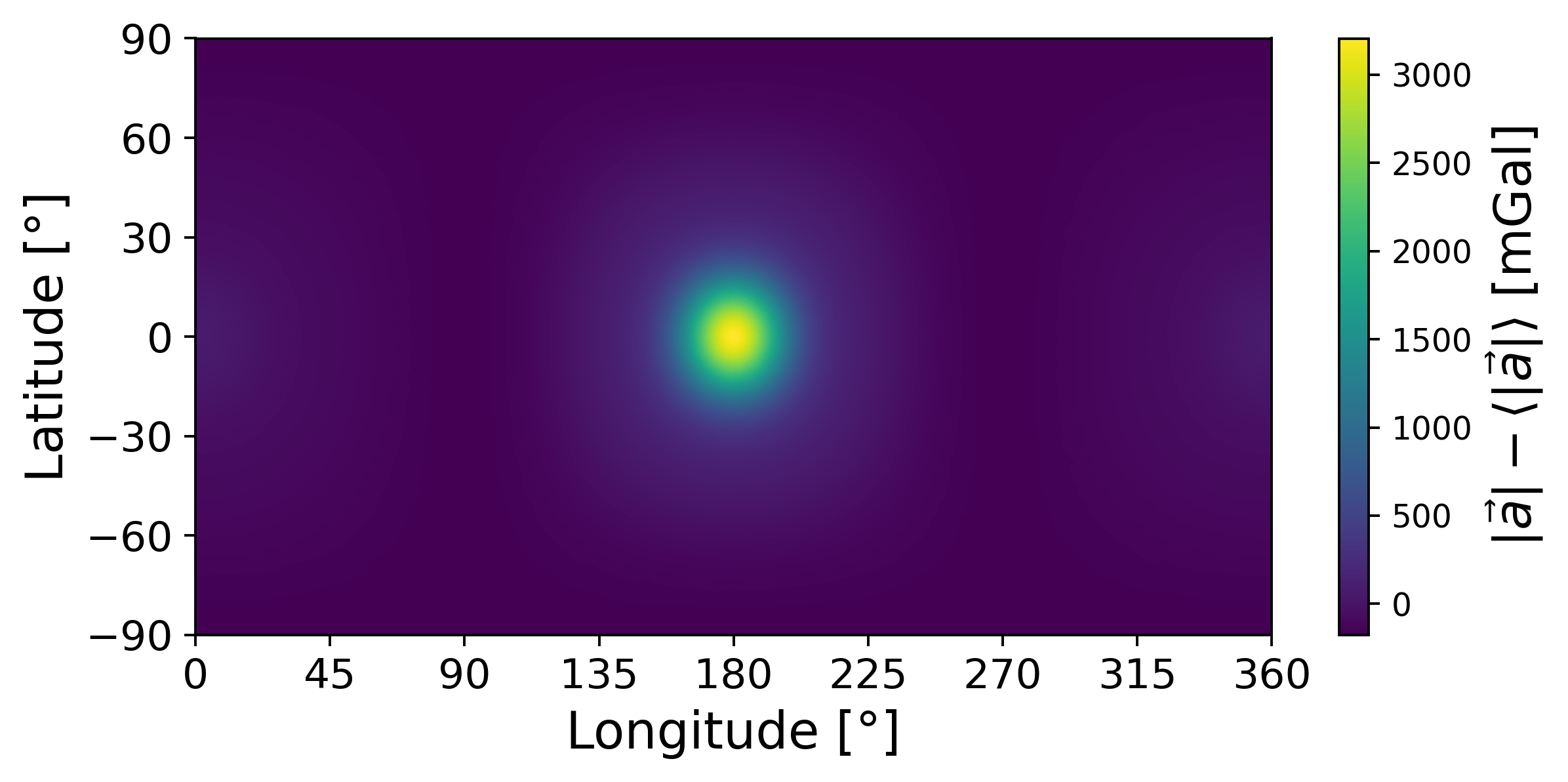}
\caption{Gravity anomaly maps of the recentered post-impact body, sampled on a HEALPix tracer shell at a fixed radius of $\SI{0.4}{\Rearth}$. Each panel shows, at each tracer location, the deviation of the local acceleration magnitude $\vert\vec{a}\vert$ from its mean over the full shell, $\vert\vec{a}\vert - \langle\vert\vec{a}\vert\rangle$, in \si{\milli\Gal}, as a function of longitude and latitude. Top: the total field, dominated by the unrelaxed basin cavity. Bottom: the field of the core material alone, showing the positive anomaly produced by the impact-induced core dome. The computation of both maps is described in Appendix~\ref{sec:gravity_method}.}
\label{fig:gravity_map}
\end{figure}

We further find that the impact measurably thins the mantle beneath the basin, and that Mercury's core, normally close to spherical, is deformed and bulges upward locally beneath the impact site. At the end of the simulation, the mantle column directly beneath the basin center is thinned from its pre-impact thickness of $\approx\SI{470}{\kilo\meter}$ to $\approx\SI{330}{\kilo\meter}$, and the core surface beneath the impact site is raised by $\approx\SI{120}{\kilo\meter}$ above the mean core radius. This time evolution is illustrated in Figure~\ref{fig:Impact_Composite_Figure}, which shows the progressive thinning of the mantle column and the development of the core protrusion as the simulation proceeds.

This finding is qualitatively consistent with the mantle-uplift mechanism invoked to explain mascons generally \citep{meloshOriginLunarMascon2013,gosselinFormationCalorisBasins2025}: a locally thinned mantle (and, in our case, an even more extreme local excess of dense core material near the surface) is precisely the kind of mass excess needed to produce a positive gravity anomaly once the transient excavation signature has relaxed away. However, we cannot determine from our simulations alone whether this deformed, domed-upward core configuration is a permanent feature or a transient one. Our simulations do not run for anywhere near the geologic timescales (thermal and viscous relaxation, lithospheric thickening, possible volcanic infilling) over which the basin is expected to evolve towards its present-day state \citep{meloshOriginLunarMascon2013,gosselinFormationCalorisBasins2025}, so we cannot rule out that the core protrusion relaxes substantially, or even entirely, on longer timescales. We therefore present it as qualitative support for the mantle/core-uplift class of hypotheses, rather than as direct quantitative confirmation of the observed anomaly.

The raw gravity signal computed directly from our simulation output is, in fact, dominated by the unrelaxed basin itself: this unrelaxed cavity produces an anomaly of several thousand \si{\milli\Gal}, well over an order of magnitude larger than the $\sim\SI{100}{\milli\Gal}$ level of the observed present-day anomaly. This masks any deeper contribution in the total field (Figure~\ref{fig:gravity_map}, top panel). The contribution of the deformed core can, however, be isolated by recomputing the tracer-shell gravity field for the core material alone (Appendix~\ref{sec:gravity_method}). Because the pre-impact core is spherically symmetric and would produce a uniform field on the tracer shell, every feature in this core-only map is impact-induced. The resulting map (Figure~\ref{fig:gravity_map}, bottom panel) shows a positive anomaly of $\approx\SI[explicit-sign = +]{+3000}{\milli\Gal}$ coincident with the basin: the direct gravitational expression of the core dome. Taken at face value, this unrelaxed deep contribution exceeds the observed anomaly by a factor of $\sim\num{30}$, so the dome cannot persist in its immediate post-impact form. It demonstrates, however, that the impact emplaces a deep, positive gravity contribution of the correct sign and location. Only a small fraction of it would need to survive the subsequent viscous and thermal evolution to account for the observed $\sim\SI{100}{\milli\Gal}$ mascon. The impact clearly produces a superisostatic deep mass excess; the question of the anomaly's origin thus shifts to what fraction of this excess survives relaxation. Our simulations cannot address this, and we identify it as the key target for coupled long-term evolution models (Section~\ref{sec:Summary_and_Conclusions}).

\subsection{The Antipodal Terrain}\label{sec:antipode}

Directly opposite the Caloris basin lies an anomalous region of hilly and lineated terrain, informally dubbed the ``weird terrain'' by the Mariner~10 team \citep{murrayMariner10Pictures1974}, that disrupts pre-existing landforms including crater rims over an area of at least \SI{5e5}{\kilo\meter\tothe2} \citep{meloshTectonicsMercury1988}. The terrain consists of closely packed hills and depressions \SIrange{5}{10}{\kilo\meter} across and \SIrange{0.1}{1.8}{\kilo\meter} high, and its age is broadly consistent with that of the Caloris basin itself \citep{meloshTectonicsMercury1988}. The terrain sits close to the antipode of one of the largest impact structures on Mercury, and analogous, if smaller, disrupted terrains are found antipodal to the Imbrium and Orientale basins on the Moon. It has therefore long been interpreted as a direct consequence of the Caloris-forming impact.

\subsubsection{Hypotheses for the origin of the antipodal terrain} \label{sec:antipode_hypotheses}

Two impact-related mechanisms dominate the literature. The first, \emph{seismic focusing}, was proposed by \citet{schultzSeismicEffectsMajor1975}, who suggested that seismic waves generated at the impact site propagate through the planet and are constructively refocused at the antipode by the spherical geometry of the body, concentrating enough energy to fracture and disrupt the surface. Early numerical estimates for the Caloris event by \citet{hughesGlobalSeismicEffects1977} suggested vertical ground motions of order one kilometer around the antipode, broadly consistent with the observed relief of the hilly and lineated terrain. More recent seismic modeling by \citet{luSeismicEffectsCaloris2011}, using physically motivated interior-structure models for Mercury, found that direct body waves are too weak to account for the disruption on their own. They instead argued for a combination of high-frequency, high-stress guided mantle waves (trapped between the core and the free surface) and antipodally focused Rayleigh waves, responsible for the crustal fracturing and for the large vertical displacements, respectively. Notably, their modeled zone of disruption remained smaller than the observed extent of the terrain, implying either that Mercury's shallow structure modulates the antipodal response in ways not captured by their models, or that the seismic energy imparted by the impact exceeded their assumed values.

The second mechanism invokes \emph{ballistic and vapor-plume convergence of ejecta}: material launched from the impact site on trajectories energetic enough to travel most of the way around the planet reconverges geometrically at the antipode. On the Moon, such antipodal convergence is both predicted by three-dimensional simulations of basin-forming impacts \citep{hoodAntipodalEffectsLunar2008} and observed for the young crater Tycho, whose antipode hosts anomalous melt ponds and rocky deposits attributed to converging ballistic ejecta \citep{robinsonExceptionalGroupingLunar2016,bandfieldDistalEjectaLunar2017,adlerTychoEjectaDeposits2019}. The simulations of \citet{hoodAntipodalEffectsLunar2008} find antipodal deposit thicknesses of order one kilometer for $\sim\SI{200}{\kilo\meter}$-diameter impactors, comparable in scale to the relief of the hilly and lineated terrain. The same convergence mechanism has also been invoked to explain antipodal magnetic anomalies on the Moon through the delivery and shock magnetization of impact-derived material \citep{hoodAntipodalEffectsLunar2008}. For Mercury itself, recent magnetohydrodynamic modeling of a Caloris-scale impact suggests that impact-generated plasma can transiently amplify the ancient magnetic field and record it as shock remanent magnetization at the basin antipode \citep{narrettImpactPlasmaAmplification2026}. Basin-scale ejecta modeling has likewise been used to link antipodal deposits to specific lunar basins \citep{wieczorekSerenitatisOriginImbrian2001}.

More recently, this impact-antipodal link itself has been questioned. \citet{rodriguezChaoticTerrainsMercury2020} re-examined the terrain, which they classify among Mercury's chaotic terrains, using higher-resolution MESSENGER imagery and altimetry. They found that its resurfacing continued until $\sim\SI{1.8}{\giga\year}$, roughly two billion years after the Caloris-forming impact, and identified morphologically similar chaotic terrains on Mercury with no antipodal basin at all. They therefore argued for a volatile-loss (sublimation-driven collapse) origin unrelated to any single impact event, rather than the classical seismic or ejecta hypotheses. We do not attempt to settle this debate here, but note it as a caveat: even a quantitatively convincing seismic or ejecta signal at the antipode does not, by itself, rule out a substantial or dominant non-impact contribution to the terrain's present-day appearance.

In the following, we present a quantitative assessment of the two classical, impact-related hypotheses, antipodal seismic shaking and antipodal deposition of impact-derived material, using our impact simulations. As in the preceding section, the simulations from which the antipodal diagnostics are extracted are vertical impacts (Section~\ref{sec:gravity}). Vertical incidence is the geometry most favorable to antipodal convergence, since both the seismic energy and the ejecta of an axisymmetric impact refocus on the exact antipode. Our antipodal estimates should therefore be read as upper bounds on the strength and symmetry of the antipodal response. This interpretation is consistent with the observed hilly and lineated terrain being centered close to, but not exactly at, the Caloris antipode \citep{murrayMariner10Pictures1974,meloshTectonicsMercury1988}.

\subsubsection{Impact-induced accelerations at the antipode}\label{sec:antipode_accel}
\begin{figure}[ht!]
\centering
\includegraphics[width=\linewidth]{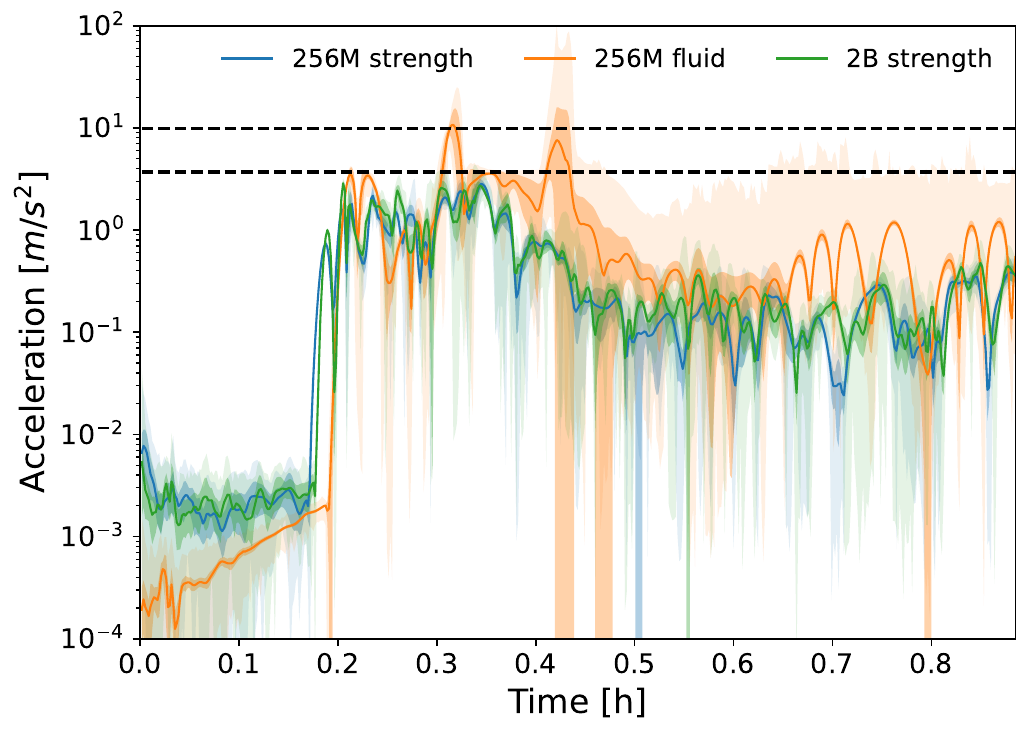}
\caption{Time series of the acceleration in a crustal cap of \SI{10}{\degree} angular radius centered on the antipode, for the 256-million-particle simulations (CM profile, \SI{72.5}{\kilo\meter} crust) with material strength enabled and disabled, and for the 2-billion-particle simulation with material strength. The solid lines show the median over the cap; the shaded regions show the 16--84 percentile range and the minimum--maximum range. The lower dashed line denotes Mercury's surface gravity, $g_{\rm M}=\SI{3.70}{\meter\per\second\tothe2}$, the threshold above which surface material is momentarily lofted; the upper dashed line denotes Earth's surface gravity, $g_\oplus=\SI{9.81}{\meter\per\second\tothe2}$, the reference acceleration commonly used in the terrestrial seismic literature.}
\label{fig:antipode_accel}
\end{figure}

To assess the seismic-shaking hypothesis, we extracted the peak particle accelerations induced by the impact within a spherical cap of \SI{10}{\degree} angular radius centered on the antipode. The time window shown in Figure~\ref{fig:antipode_accel} (the first \SI{53}{\minute} after impact) ends well before the first impact-derived material reaches the antipodal region; over this window, the cap contains only the in-place antipodal crust and the measured accelerations are purely seismic in origin. The physically relevant threshold for shaking-induced mobilization of surface material is the local gravitational acceleration: ground accelerations exceeding Mercury's surface gravity, $g_{\rm M}=\SI{3.70}{\meter\per\second\tothe2}$, momentarily unload the surface and can loft loose material ballistically. This is the criterion commonly applied to impact-induced seismic shaking on small bodies \citep{richardsonImpactinducedSeismicActivity2004}. For comparison with the terrestrial seismic literature, we also indicate Earth's surface gravity, $g_\oplus=\SI{9.81}{\meter\per\second\tothe2}$. Figure~\ref{fig:antipode_accel} shows the resulting accelerations for the 256-million-particle simulations run with material strength enabled and disabled, as well as for the 2-billion-particle simulation with material strength.

When strength is included, the median acceleration in the cap remains well below even Mercury's surface gravity $g_{\rm M}$ throughout the simulation; only the extreme of the minimum--maximum envelope exceeds $g_{\rm M}$, marginally and on a single occasion, and all accelerations remain well below $g_\oplus$. At most, then, isolated particles within the cap are briefly and marginally unloaded, while the cap as a whole never approaches the mobilization threshold. Taken at face value, this is difficult to reconcile with a purely seismic-shaking origin for the hilly and lineated terrain. Shaking of this magnitude is not obviously sufficient to mobilize kilometer-scale hills and lineations, or to fracture pre-existing crustal material outright. It could trigger large-scale failure only if the antipodal crust was already substantially weakened, for example thermally or by a pre-existing zone of structural heterogeneity. This would be consistent with the guided-wave picture of \citet{luSeismicEffectsCaloris2011}, in which the relevant driver is not a single large acceleration pulse but the cumulative, resonant build-up of trapped mantle and surface waves over the course of the seismic ringing. A peak-acceleration diagnostic is not designed to capture such an effect; we probe this cumulative regime directly in Section~\ref{sec:antipode_strain}.

When strength is switched off, we instead find two pronounced spikes in the antipodal acceleration well in excess of both $g_{\rm M}$ and $g_\oplus$. It is tempting to associate these excursions with the onset of disruptive shaking, but we treat them with caution. Removing material strength entirely is not physical: it eliminates all shear resistance in the crust and mantle, so the strengthless case should be understood as a bounding, limiting-case scenario rather than a realistic prediction. The two spikes likely correspond to the arrival of distinct focused wave phases at the antipode (e.g., a direct/refracted body-wave arrival followed by a later surface- or guided-wave arrival), consistent with the multi-phase picture of \citet{luSeismicEffectsCaloris2011}. Their amplitude, however, is almost certainly overestimated by the absence of strength-related damping and yield. We therefore regard the true antipodal accelerations as bounded between our two limiting cases: too weak to drive disruption on their own in the strength-included case, and almost certainly overestimated in the strengthless case. Once local weakening, fracture-induced strength loss, and resonant wave trapping are accounted for self-consistently, the physical answer likely lies closer to the lower end.

\subsubsection{Accumulated strain at the antipode}\label{sec:antipode_strain}
\begin{figure}[ht!]
\centering
\includegraphics[width=\linewidth]{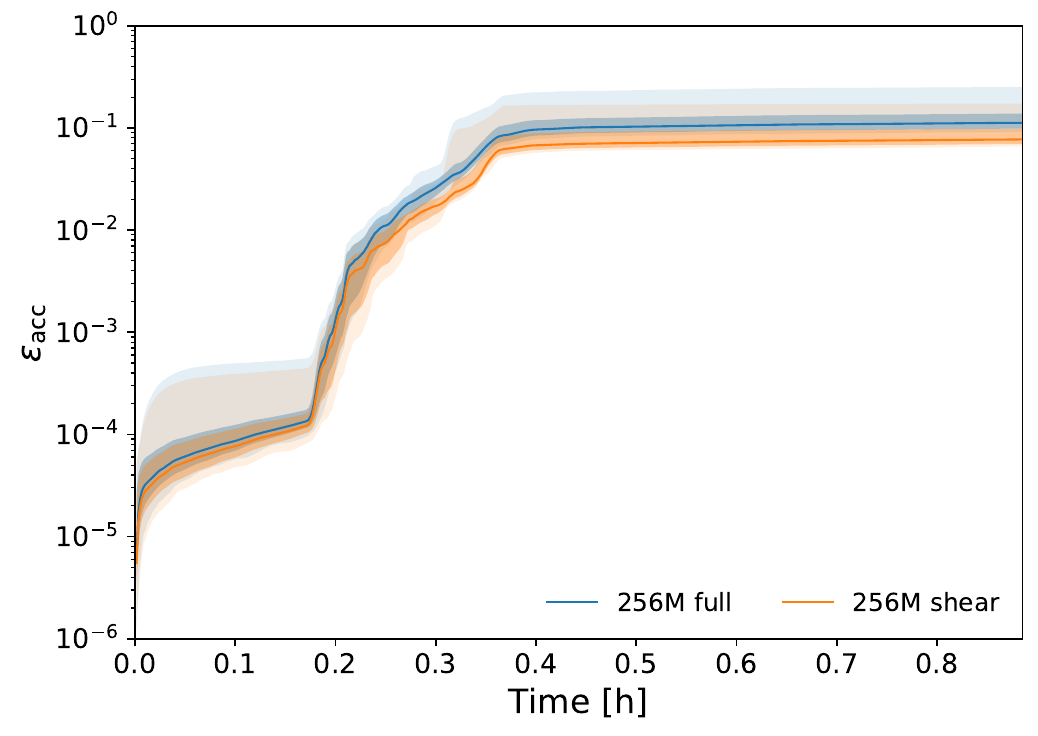}
\caption{Time series of the accumulated strain $\epsilon_{\rm acc}$ (Equation~\eqref{eq:accumulated_strain}) in the crustal cap of \SI{10}{\degree} angular radius centered on the antipode, for the 256-million-particle simulation (CM profile, \SI{72.5}{\kilo\meter} crust) with material strength. The two curves show the two integrand variants: the full strain-rate tensor and its deviatoric (shear-only) part. The solid lines show the median over the cap; the shaded regions show the 16--84 percentile range and the minimum--maximum range.}
\label{fig:antipode_strain}
\end{figure}

The peak-acceleration diagnostic of the previous section asks whether any single wave arrival is strong enough to unload the antipodal surface. The cumulative picture of \citet{luSeismicEffectsCaloris2011} instead attributes the damage to the deformation built up over the entire seismic wave train. To probe this regime directly, we repeated the 256-million-particle strength simulation (CM profile, \SI{72.5}{\kilo\meter} crust) with a per-particle accumulated-strain diagnostic enabled; the run is otherwise identical (Section~\ref{sec:Initial_conditions}). For every particle, the code integrates the accumulated strain

\begin{equation}
\epsilon_{\rm acc}(t)=\int_0^t \sqrt{J_2\left(\dot{\epsilon}\right)}\, dt'\,,
\label{eq:accumulated_strain}
\end{equation}

\noindent where $\dot{\epsilon}^{\alpha\beta}=\frac{1}{2}\left(\partial^\alpha v^\beta + \partial^\beta v^\alpha\right)$ is the strain-rate tensor, evaluated for each particle from the kernel-smoothed velocity gradients; the discretized SPH form is given in \citet{meierSmoothedParticleHydrodynamics2026a}. $J_2$ denotes its second invariant. We track two variants of this quantity. The first uses the full strain-rate tensor and therefore accumulates all deformation, including the volumetric compression and dilatation carried by compressional wave arrivals. The second uses only the deviatoric part, $s^{\alpha\beta}=\dot{\epsilon}^{\alpha\beta}-\frac{1}{3}\dot{\epsilon}^{\gamma\gamma}\delta^{\alpha\beta}$, and thus isolates the shear deformation relevant to material failure. Both are quoted in the $\sqrt{J_2}$ convention; the von Mises equivalent strain is larger by a constant factor of $2/\sqrt{3}$. One property of this measure is essential for its interpretation: it accumulates all deformation, including fully reversible elastic oscillation, and is therefore an upper bound on the permanent strain.

Figure~\ref{fig:antipode_strain} shows the accumulated strain in the same antipodal cap as in Figure~\ref{fig:antipode_accel}; the time window again ends before the first impact-derived material reaches the antipodal region, so the deformation is purely seismic in origin. Before the first seismic arrival, the accumulated strain grows only slowly and reaches $\sim\num{1.5e-4}$. This baseline is dominated by residual particle noise in the SPH velocity-gradient estimates and sets the noise floor of the diagnostic; the seismic signal that follows rises two to three orders of magnitude above it. The first sharp rise begins $\approx\SI{11}{\minute}$ after impact, consistent with the transit time of the direct compressional wave through Mercury and coincident with the onset of large accelerations in Figure~\ref{fig:antipode_accel}. The rise continues through a sequence of smaller steps until $\approx\SI{22}{\minute}$ after impact, reflecting the successive arrival of slower and multiply reflected wave phases, and then settles onto a plateau. The median accumulated strain at the plateau is $\approx\num{0.11}$ (16--84 percentile range \numrange{0.10}{0.14}), with a cap minimum of $\approx\num{0.09}$ and a maximum of $\approx\num{0.25}$. The first arrival contributes only $\sim\num{5e-3}$ of this total. The bulk of the deformation is thus built up by the later, sustained sequence of arrivals, in line with the dominance of guided and surface waves over direct body waves found by \citet{luSeismicEffectsCaloris2011}. The deviatoric variant tracks the full-tensor curve through the entire wave train and plateaus at a median of $\approx\num{0.08}$ (16--84 percentile range \numrange{0.07}{0.10}; cap minimum \num{0.066}), about \SI{30}{\percent} below the full-tensor value. Most of the accumulated deformation is therefore shear rather than volumetric compression.

For the interpretation of the plateau value, the natural comparison is the elastic strain at yield in our strength model, $Y/(2\Gamma)$. With the parameters of Table~\ref{tab:Strength_parameters}, this ratio is at most $Y_{\rm m}/(2\Gamma)\approx\num{2e-2}$, reached only at high confining pressure; at the low pressures of the near-surface, the frictional yield strength and hence the yield strain are far smaller. Even the shear-only measure exceeds the maximum yield strain by a factor of a few, and the near-surface yield strain by orders of magnitude. Because the measure includes reversible elastic cycling, this does not imply a permanent deformation of ten percent. It does imply that the antipodal crust is cyclically driven, over roughly ten minutes of seismic ringing, through a total deformation far larger than it can accommodate elastically, so a substantial part of it must be taken up by frictional sliding and plastic flow. The two seismic diagnostics therefore refine, rather than contradict, each other. No single wave arrival unloads or shatters the antipodal surface, but the deformation accumulated over the full wave train is more than sufficient to progressively weaken, fracture, and rework the antipodal crust. This is precisely the cumulative regime advocated by \citet{luSeismicEffectsCaloris2011}. It identifies the seismic contribution to the antipodal terrain as one of progressive weakening rather than a single disruptive pulse, and it supplies a concrete mechanism for the preconditioning discussed in Section~\ref{sec:antipode_synthesis}: the material converging on the antipode (Section~\ref{sec:antipode_mass}) arrives at a surface that the seismic wave train has already deformed well past its elastic limit.

\subsubsection{Mass and equivalent deposit thickness at the antipode}\label{sec:antipode_mass}

\begin{table*}[t]
\centering
\begin{tabular}{lcccccc}
\toprule
Resolution & Crust thickness & $t$ [\si{\hour}] & $M_{{\rm tar,m}}$ [\si{\kilo\gram}] & $M_{{\rm tar,c}}$ [\si{\kilo\gram}] & $M_{\rm imp}$ [\si{\kilo\gram}] & $h_{\rm eq}$ [\si{\kilo\meter}] \\
\midrule
32M   & \SI{104}{\kilo\meter}  & 3.54 & \num{2.3e17} (23)   & \num{1.6e18} (169)  & \num{1.5e17} (14)   & 1.33 \\
256M  & \SI{72.5}{\kilo\meter} & 3.54 & \num{5.1e17} (393)  & \num{6.2e17} (451)  & \num{5.0e17} (388)  & 1.08 \\
256M  & \SI{104}{\kilo\meter}  & 1.77 & \num{3.0e16} (23)   & \num{8.3e17} (616)  & \num{1.2e18} (900)  & 1.35 \\
2048M & \SI{40}{\kilo\meter}   & 1.77 & \num{6.7e17} (4176) & \num{6.6e17} (3929) & \num{2.7e18} (16304) & 2.68 \\
\bottomrule
\end{tabular}
\caption{Material originating from the impact-side hemisphere that has reached the antipodal cone (selection criteria in the text), for the series of runs varying resolution and crustal thickness, at $v=\SI{21}{\kilo\meter\per\second}$, $\theta=\SI{0}{\degree}$, $R=\SI{150}{\kilo\meter}$, all using the CM thermal profile, measured at time $t$ after impact. The masses are split into target mantle ($M_{{\rm tar,m}}$), target crust ($M_{{\rm tar,c}}$), and impactor ($M_{\rm imp}$) material, with the number of particles behind each entry given in parentheses; the last column gives the equivalent layer thickness $h_{\rm eq}$ for homogeneous deposition over the antipodal cap at \SI{2.65}{\gram\per\centi\meter\tothe3}.}
\label{tab:antipode_mass}
\end{table*}

To assess the ejecta-convergence hypothesis, we measured the mass of material originating from the impact-side hemisphere that has reached the antipodal region, in the four runs varying resolution and crustal thickness (all CM profile; Table~\ref{tab:antipode_mass}). Specifically, we count every particle that (i) lies within a cone of \SI{10}{\degree} angular radius (half-opening angle) centered on the antipode direction, (ii) is located at a radius $r>\SI{0.2}{\Rearth}$, and (iii) originated in the impact-side hemisphere of the pre-impact configuration. Criterion (iii) is evaluated by mapping each particle back to its position in the initial conditions via its unique particle ID; it restricts the census to material transported from the impact-side hemisphere to the antipode, which in practice arrives at or above the surface, ballistically or as vapor. Criterion (ii) serves only to exclude core material: because the planet as a whole is displaced slightly during the simulation, deep-interior particles, which also originate on the impact side, would otherwise move into the antipodal cone direction and enter the selection. We make no distinction between vaporized, molten, and solid material. The census is taken at the final snapshot of each run (\SI{1.77}{} or \SI{3.54}{\hour} after impact; Table~\ref{tab:antipode_mass}). Unlike the basin diameter, the antipodal mass budget is not stationary over this interval, since convergence may still be ongoing; the entries at different times are therefore not directly comparable. Table~\ref{tab:antipode_mass} reports the masses by source (impactor, crust, mantle), the number of particles behind each entry, and the equivalent layer thickness $h_{\rm eq}$ that would result if the total mass were spread homogeneously over the antipodal cap at a bulk density of \SI{2.65}{\gram\per\centi\meter\tothe3}, representative of crustal material.

The equivalent thickness is $h_{\rm eq}\approx\SIrange{1.1}{1.35}{\kilo\meter}$ in the three lower-resolution runs and $h_{\rm eq}\approx\SI{2.7}{\kilo\meter}$ in the 2-billion-particle run. We regard the latter as the most reliable estimate: it is the only run in which all three source components are resolved by thousands of particles, whereas several entries of the lower-resolution runs correspond to only \numrange{14}{23} particles and are thus subject to \SIrange{20}{27}{\percent} counting noise (parenthetical counts in Table~\ref{tab:antipode_mass}). The composition of the antipodal material is, however, clearly not converged across the series: the impactor-derived mass in the cone grows systematically with resolution, from $\sim\SI{1.5e17}{\kilo\gram}$ at 32M to $\sim\SI{2.7e18}{\kilo\gram}$ at 2B particles. This trend is too large to be explained by counting noise. It plausibly reflects the better-resolved fragmentation and vapor-plume expansion of the impactor at higher resolution, compounded by the differing crust thicknesses and measurement times across the series. We therefore quote the antipodal deposit as an equivalent thickness of \SIrange{1}{3}{\kilo\meter}, with the upper half of this range favored by the best-resolved run, and defer a converged compositional breakdown to future work. If a substantial fraction of the antipodal deposit is indeed impactor-derived, as the best-resolved run suggests, it could carry a compositional, and hence spectral, signature distinct from the surrounding crust, providing a potential observational discriminant for the ejecta-convergence hypothesis. This thickness is directly comparable to the antipodal ejecta thicknesses of order \SI{1}{\kilo\meter} found in three-dimensional simulations of large lunar basin-forming impacts \citep{hoodAntipodalEffectsLunar2008}, and to the kilometer-scale relief of the hilly and lineated terrain itself \citep{meloshTectonicsMercury1988}. This lends independent, if approximate, support to the ejecta-convergence hypothesis: a substantial mass of impactor- and target-derived material is delivered to the antipodal region in our simulations, at a thickness sufficient in principle to account for the observed terrain relief, either directly as a deposited layer or indirectly through the disruption caused by the high-velocity arrival of this material.

We emphasize, however, that this estimate is subject to several important caveats. First, whether this converging material actually reaches the surface as coherent ballistic ejecta, as an impact-vapor plume, or as some mixture of both is difficult to establish robustly in our simulations. The thermodynamic state of material at these extreme shock pressures is sensitive to the equation of state. Critically, our estimate of which particles are vaporized versus condensed or solid is also influenced by the interface and free-surface correction scheme used in the code, which is known to affect the computed pressure and internal-energy states of particles near a free surface. A vapor-dominated versus solid-dominated convergence would produce qualitatively different surface expressions (a thin, smoothly draped condensate blanket versus discrete, blocky hummocky deposits), and we cannot yet distinguish between these outcomes with confidence. Second, the assumption of a homogeneous deposit over the cap is a simplification adopted purely to obtain a characteristic thickness scale for comparison with observations. In reality, both ballistic and vapor-plume deposition are expected to be laterally non-uniform; the true deposit would be patchier, and plausibly thicker in some sub-regions, than the cap-averaged value reported here. Third, our reported thickness is an instantaneous, post-impact estimate. It does not account for subsequent viscous relaxation, volcanic embayment, or long-term erosion and degradation of the type invoked by \citet{rodriguezChaoticTerrainsMercury2020}, any of which could substantially modify the present-day expression of such a deposit over the intervening $\sim\SI{3.9}{\giga\year}$. In the same vein, the equivalent thicknesses of Table~\ref{tab:antipode_mass} are computed at the solid crustal density of \SI{2.65}{\gram\per\centi\meter\tothe3}. If the converging material instead arrives as vapor and recondenses as a porous blanket rather than as competent rock, the resulting deposit would be correspondingly thicker, so the quoted values are conservative in that respect. Conversely, a true crustal density above the adopted \SI{2.65}{\gram\per\centi\meter\tothe3} ($\approx\SIrange{2.7}{3.1}{\gram\per\centi\meter\tothe3}$; Section~\ref{sec:Models}) would proportionally reduce the quoted thicknesses.

\subsubsection{Relative importance of the two mechanisms}\label{sec:antipode_synthesis}

Taken together, our results do not single out either the seismic-shaking or the ejecta-convergence mechanism as solely responsible for the hilly and lineated terrain, and are compatible with a picture in which both contribute in combination, as has been suggested for antipodal terrains on other bodies \citep{schultzSeismicEffectsMajor1975,wieczorekSerenitatisOriginImbrian2001,hoodAntipodalEffectsLunar2008}. In the physically more defensible strength-included case, the antipodal accelerations we find are arguably too low to drive disruption in isolation; the strengthless case that does exceed $g_{\rm M}$ should be treated as an upper bound rather than a physical prediction. The accumulated-strain diagnostic (Section~\ref{sec:antipode_strain}) sharpens this picture: although no single arrival is disruptive, the deformation built up over the full wave train exceeds the elastic limit of the crust by a large margin, so the seismic contribution plausibly acts through cumulative weakening rather than through direct, instantaneous disruption. In contrast, the mass of impact-derived material reaching the antipodal region is large enough, at face value, to produce a kilometer-scale deposit consistent with the observed terrain relief, though this result carries substantial uncertainty tied to the treatment of vaporization and the free-surface correction in our simulations.

Our simulations moreover show that these two contributions are sequential rather than simultaneous: the seismic pulse arrives and decays before the converging material does (Section~\ref{sec:antipode_accel}). The accumulated strain shows that this ordering matters: the seismic wave train deforms the antipodal surface well past its elastic limit before the converging material arrives, fracturing and loosening it for the subsequent deposition, making the combined effect potentially larger than either mechanism alone. Under the oblique incidence favored by the basin-size analysis, both the seismic focusing and the ejecta convergence would weaken and shift in tandem, so the relative ranking of the two mechanisms is likely more robust than either absolute amplitude. A more definitive assessment would require simulations with a higher-fidelity, damage-coupled strength model capable of capturing the resonant, multi-phase wave trapping identified by \citet{luSeismicEffectsCaloris2011}, together with an equation-of-state and near-surface treatment validated specifically for the vapor/melt/solid partitioning of antipodally converging material.

\section{Summary and Conclusions}\label{sec:Summary_and_Conclusions}

To our knowledge, this is the first study to address basin size, the Caloris gravity anomaly, and the antipodal effects of the impact within a single, self-consistent, fully three-dimensional simulation framework. The basin-size survey covers oblique impact geometries, while the gravity and antipodal analyses are performed for vertical incidence. At a resolution of 32 million particles, we surveyed all combinations of three impactor radii (\SIrange{100}{200}{\kilo\meter}), five impact velocities (\SIrange{12}{48}{\kilo\meter\per\second}), five impact angles (\SIrange{0}{60}{\degree} from vertical), and three target thermal profiles, for a total of 225 simulations. This survey was complemented by a series of runs varying resolution and crustal thickness at fixed impact conditions, with up to two billion particles; at that resolution, the observationally motivated \SI{40}{\kilo\meter} crust is resolved by ten particle layers. At all resolutions, the basin dimensions are measured directly from the crust particles (Appendix~\ref{sec:appendix:Measuring_Crater_Size}).

Across the parameter survey, the effective basin diameter ranges from $\approx\SI{550}{\kilo\meter}$ to $\approx\SI{4000}{\kilo\meter}$ and is well described by the power law $D_{\rm eff} \propto R^{1.14\pm0.02}\,v^{0.66\pm0.01}\,\cos(\theta)^{0.56\pm0.03}$. The fitted amplitude is $D_0 = \SI{1509\pm15}{\kilo\meter}$, evaluated at the reference conditions ($R=\SI{150}{\kilo\meter}$, $v=\SI{21}{\kilo\meter\per\second}$, vertical incidence), and the three thermal profiles individually return very similar exponents. The target's thermal structure therefore mainly rescales the overall basin size rather than changing its functional dependence on the impactor properties. All three fitted exponents are systematically steeper than the predictions of idealized point-source scaling for competent rock, which we attribute to the basin-scale, thermally structured nature of the cratering process, well outside the regime of simple bowl-shaped craters in which the literature scaling laws are calibrated. The ratio of the velocity to the radius exponent nevertheless corresponds to a coupling exponent $\mu = \num{0.58\pm0.01}$, within the dimensionally allowed range and close to literature values for competent rock; the deviation from point-source scaling is localized in the overall power with which the coupling parameter enters. We further find that the angle dependence is not strictly monotonic, with the effective diameter frequently peaking at intermediate obliquity due to basin elongation. Energy-only scaling describes the vertical-incidence data well, where the effective coupling exponent is consistent with pure energy coupling, but fails once oblique impacts are included.

The sensitivity runs show that neither thinning the crust at fixed resolution nor increasing the resolution at fixed crust thickness produces a trend that is consistent in sign across the three thermal profiles; the individual effects are of order \SIrange{4}{17}{\percent} in $D_{\rm eff}$, with the cold-mantle profile responding with the opposite sign to the other two profiles in both comparisons. The single simulation at the highest resolution and the observationally motivated crust thickness of \SI{40}{\kilo\meter} (two billion particles) yields a basin \SI{23}{\percent} larger than the corresponding 32-million-particle baseline, although the resolution and crust-thickness contributions to this difference cannot currently be separated. Basin sizes measured from simulations of this type should therefore be understood to carry a systematic uncertainty of this order, and dedicated series of runs varying resolution and crustal thickness one at a time are required before formal convergence can be claimed.

Comparing to the observed size of the Caloris basin (rim-to-rim diameter of approximately \SI{1550}{\kilo\meter}; area-equivalent diameter $D_{\rm eff}\approx\SI{1420}{\kilo\meter}$), we find that the observed basin size is reproduced by a broad family of impact conditions rather than a unique combination. At the survey conditions, \SI{150}{\kilo\meter} impactors at \SIrange{21}{30}{\kilo\meter\per\second} span $D_{\rm eff}\approx\SIrange{1020}{2040}{\kilo\meter}$ depending on impact angle and thermal profile, matching the observed value at oblique incidence ($\theta\approx\SIrange{42}{57}{\degree}$), a geometry independently suggested by the ellipticity of the observed basin rim. Correcting for the resolution- and crust-thickness systematics shifts this family toward smaller or slower impactors, by up to a factor of $\approx0.83$ in radius or $\approx0.73$ in velocity if the full \SI{23}{\percent} correction of the physical-crust run applies. The basin diameter alone thus does not uniquely determine the Caloris-forming impactor; instead, our scaling relations map the observed basin size onto an admissible region of impactor parameter space, with a quantified systematic uncertainty attached.

Regarding the origin of the Caloris gravity anomaly, we find that all impactor material remaining in the basin region is vaporized. No buried remnant therefore exists to source the anomaly, arguing against the classical buried-impactor hypothesis for crust-like impactors. Our undifferentiated impactors cannot, however, address whether the dense metallic core of a differentiated impactor, which is more resistant to vaporization than the rocky material simulated here, could survive at depth. Instead, the impact measurably thins the mantle beneath the basin and raises a local dome on Mercury's otherwise nearly spherical core, precisely the type of superisostatic mass excess invoked by the mantle-uplift and thermal-contraction models for lunar and Mercurian mascons. Although the total simulated gravity field is dominated by the immediate post-impact, unrelaxed basin cavity, isolating the field of the core material alone reveals a positive anomaly of $\approx\SI[explicit-sign = +]{+3000}{\milli\Gal}$ coincident with the basin: the direct gravitational expression of the core dome, and roughly 30 times the observed anomaly. Because our simulations cannot follow the subsequent thermal and viscous evolution over geologic timescales, the surviving fraction of this deep contribution remains open; only a small fraction would need to persist to account for the observed mascon.

At the antipode, we quantified both classical impact-related mechanisms proposed for the hilly and lineated terrain. With material strength included, the median acceleration in a cap of \SI{10}{\degree} angular radius around the antipode remains well below Mercury's surface gravity, $g_{\rm M}=\SI{3.70}{\meter\per\second\tothe2}$, the threshold for lofting surface material; only the extreme of the minimum--maximum envelope marginally exceeds $g_{\rm M}$ on a single occasion. The seismic shaking is thus insufficient on its own to disrupt intact crust, while the strengthless limiting case produces spikes well above $g_{\rm M}$ that overestimate the true response. The accumulated strain in the same cap nevertheless reaches $\approx\num{0.1}$ ($\approx\num{0.08}$ for its shear-only part) by the end of the seismic ringing, far above the elastic limit of the crust, indicating that the seismic contribution acts through cumulative deformation over many wave passages rather than through any single pulse. In contrast, an equivalent thickness of \SIrange{1}{3}{\kilo\meter} of impactor- and target-derived material converges on the antipodal region, with the best-resolved simulation favoring the upper half of this range, comparable to the observed relief of the terrain. Within the caveats attached to the vapor/melt/solid partitioning of this material, our results therefore suggest that the deposition and high-velocity arrival of converging impact-derived material contributed at least as much as seismic shaking to the formation of the antipodal terrain. A combination of both mechanisms remains the most plausible picture; notably, the two act sequentially, with the seismic pulse preceding the arrival of converging material, and the accumulated strain indicates that the shaking itself can supply the weakening on which the later deposition acts. Both antipodal diagnostics are derived from vertical impacts, the geometry most favorable to antipodal convergence, and thus represent upper bounds. Beyond the two mechanical signatures quantified here, the Caloris antipode may also carry a magnetic one: impact-generated plasma can transiently amplify the ancient magnetic field and record it as crustal magnetization at the antipode \citep{narrettImpactPlasmaAmplification2026}, offering a complementary observable for BepiColombo.

Several extensions of this work are planned. Simulations with differentiated, iron-cored impactors are required to directly test whether a surviving impactor core could contribute to the gravity anomaly. A damage-coupled strength model would allow the resonant, multi-phase seismic wave trapping identified by \citet{luSeismicEffectsCaloris2011} to be captured self-consistently, sharpening both the antipodal acceleration and the accumulated-strain constraints. Measuring the post-collapse crustal structure of our simulated basins would enable a direct comparison with crustal-thickness models of the Caloris region and with the megabasin architecture identified by \citet{gosselinCrustalBlockMuted2023}, providing a basin-size-independent test of the impact conditions and of the thermal-state dependence discussed in Section~\ref{sec:caloris_implications}. Finally, coupling the post-impact state of our simulations to long-term thermal and viscous evolution models would connect the immediate impact outcome, including the mantle thinning and core dome found here, to the present-day, relaxed basin structure. This would enable a direct, quantitative comparison with the MESSENGER gravity field and with the improved gravity and altimetry data expected from the BepiColombo mission, in particular for the surviving fraction of the $\approx\SI[explicit-sign = +]{+3000}{\milli\Gal}$ core-dome contribution identified here.

\begin{acknowledgments}
This work has been carried out within the framework of the National Centre of Competence in Research PlanetS supported by the Swiss National Science Foundation under grants 51NF40\_182901 and 51NF40\_205606. The authors acknowledge the financial support of the SNSF. T.M. acknowledges support from the University of Zurich through a Candoc grant. T.M., C.R. and M.J. acknowledge support from the Swiss National Science Foundation (project numbers 200021\_207359 and 200021\_228420). We acknowledge access to Eiger.Alps at the Swiss National Supercomputing Centre, Switzerland under the University of Zurich's share with the project ID UZH4. This work was supported by a grant from the Swiss National Supercomputing Centre (CSCS) under project IDs S1285, LP87 and LP162 on Piz Daint and Daint.Alps. Parts of this manuscript were prepared with the assistance of the large language model Claude (Anthropic; Fable 5). The model was used to edit and restructure the text, to draft text passages and analysis and plotting scripts, and to perform internal consistency checks of the manuscript. All of this took place under the direction of the authors, who specified each task and its inputs; all model output was reviewed and edited by the authors and verified against the simulation data, fit results, and cited literature. The authors take full responsibility for the entire content of this publication.
\end{acknowledgments}

\begin{contribution}

T.M. performed the simulations and the analysis, created the figures, and was responsible for writing and submitting the manuscript. C.R. conceived the initial idea that led to this project, provided expertise on planetary-scale impact simulations, secured the computational resources for the simulations, and edited the manuscript. M.J. provided expertise on the material strength implementation and basin-scale cratering and edited the manuscript. J.S. is the original developer of \texttt{pkdgrav1/2/3}, obtained the funding that supported T.M., and edited the manuscript.


\end{contribution}

\section*{Data Availability}
The effective basin diameters and fitted ellipse parameters for all simulations, together with the analysis scripts used to compute the crustal-thickness maps and basin ellipse fits, the scaling-law fits, and the post-impact gravity anomaly maps, as well as the scripts generating the corresponding figures, are available at \dataset[DOI: 10.5281/zenodo.22093307]{https://doi.org/10.5281/zenodo.22093307}. The repository also includes a single 32-million-particle example snapshot, sufficient to run the complete basin-size and gravity pipelines; the full set of simulation snapshots is not archived long-term owing to its volume (up to several terabytes per run). The parameter and input files required to regenerate the simulations with \texttt{pkdgrav3}, as well as derived data products beyond those archived, are available from the corresponding author on reasonable request. \texttt{pkdgrav3} is publicly available at \href{https://bitbucket.org/dpotter/pkdgrav3}{bitbucket.org/dpotter/pkdgrav3}.

%
\facilities{Swiss National Supercomputing Centre (Piz Daint, Eiger.Alps, Daint.Alps)}

\software{pkdgrav3 \citep{potterPKDGRAV3TrillionParticle2017,meierSmoothedParticleHydrodynamics2026,meierSmoothedParticleHydrodynamics2026a},
          ballic \citep{reinhardtNumericalAspectsGiant2017},
          eoslib \citep{meierEOSlib2021,meierANEOSmaterial2021},
          tipsy \citep{n-bodyshopTIPSYCodeDisplay2011},
          numpy \citep{harrisArrayProgrammingNumPy2020},
          scipy \citep{virtanenSciPy10Fundamental2020},
          matplotlib \citep{hunterMatplotlib2DGraphics2007},
          healpy \citep{zoncaHealpyEqualArea2019},
          scikit-image \citep{waltScikitimageImageProcessing2014},
          h5py \citep{colletteH5py2024},
          GNU parallel \citep{tangeGNUParallelCommandline2011}
          }


\appendix
\section{Determination of Basin Dimensions from SPH Particle Data}\label{sec:appendix:Measuring_Crater_Size}
This appendix describes how the crustal-thickness map and the resulting basin dimensions were derived from the SPH particle output of the impact simulation. Because our simulations have sufficient resolution to model the crust as multiple particle layers, we can use the crust particles themselves to directly measure the basin size, rather than relying on a lower-resolution proxy. The procedure has three steps: constructing a crustal-thickness map on the sphere, projecting the basin outline onto a local tangent plane, and fitting an ellipse that yields the physical basin dimensions.

The simulation output is read from the snapshot, providing for each particle its mass, position $(x,y,z)$, density $\rho$, and a material identifier. Prior to any geometric analysis, the particle distribution is recentered on the center of mass of the post-impact body. To avoid biasing this estimate with distant ejecta or escaping debris, only particles within a fixed radius $r_{\rm cut}=\SI{0.395}{\Rearth}$ of the current origin are used at each iteration. The cutoff is chosen slightly larger than Mercury's radius of \SI{0.383}{\Rearth}, so that the full post-impact body is retained while distant ejecta are excluded:

\begin{equation}
\vec{r}_{\rm com} = \frac{\sum_{i:\,r_i<r_{\rm cut}} m_i\,\vec{r}_i}{\sum_{i:\,r_i<r_{\rm cut}} m_i}\,.
\end{equation}

\noindent This is subtracted from all particle positions and repeated until $\vert\vec{r}_{\rm com}\vert<\SI{5e-10}{\Rearth}$. Particles outside $r_{\rm cut}$ are then discarded as unbound or far-field material, and the remaining positions are converted from code units to km.

The recentered, retained particles are converted to spherical coordinates, colatitude $\theta_i=\arccos(z_i/r_i)$ and longitude $\phi_i=\operatorname{atan2}(y_i,x_i) \bmod 2\pi$, and assigned to a HEALPix pixel \citep{gorskiHEALPixFrameworkHighResolution2005} using the standard \texttt{ang2pix} pixelization. The resolution is chosen to match the particle count of the simulation: $N_{\rm side}=$ 128, 256, and 512, corresponding to $N_{\rm pix}=$ \SI{196608}{}, \SI{786432}{}, and \SI{3145728}{} equal-area pixels for the 32M-, 256M-, and 2B-particle runs, respectively. Crust particles are identified by their material identifier; a density cut, $\rho > \SI{2.55}{\gram\per\centi\meter\tothe3}$, additionally removes vaporized or strongly rarefied crust material, whose extended spatial distribution would otherwise contaminate the thickness estimate. For each HEALPix pixel $p$ containing at least one crust particle, the local crustal thickness is the radial extent spanned by the crust particles in that pixel,

\begin{equation}
t_p = \max_{i\in p}\, r_i-\min_{i\in p}\, r_i\,,
\end{equation}

\noindent i.e., the difference between the outermost (surface) and innermost (crust--mantle boundary) radial distance sampled in that direction. Pixels with no crust particles are assigned $t_p=0$, giving the crustal-thickness map $t_p$ shown in Figure~\ref{fig:crater_map}.

A pixel is classified as part of the basin if its thickness falls below half the global median thickness $\tilde{t}$ (the median of $t_p$ over all pixels), i.e., if $t_p < \tilde{t}/2$. This binary classification is noisy near the rim, so we clean it in two steps. First, treating pixels as nodes connected via their HEALPix neighbors (up to eight per pixel), a breadth-first connected-component search retains only the largest connected group, discarding disconnected noise elsewhere on the sphere. Second, one iteration each of morphological opening (erosion then dilation) and closing (dilation then erosion) removes small spurious rim protrusions and fills small interior holes, without altering the overall basin shape; the largest connected component is then re-selected in case the opening split the mask. The half-median threshold is robust to the presence of the basin itself: even for the largest basins in our survey ($D_{\rm eff}\approx\SI{4000}{\kilo\meter}$), the basin covers a minority of the sphere, so the global median remains anchored by the undisturbed far-field crust.

The basin dimensions are then measured by fitting an ellipse to the rim pixels (basin pixels with at least one non-basin neighbor), which first requires projecting them onto a plane. Rather than assuming the basin is centered on the HEALPix pole, we compute its true center direction $\hat{\vec{n}}$ as the normalized mean of the unit position vectors of all basin pixels,

\begin{equation}
\hat{\vec{n}} = \frac{\sum_i \hat{\vec{r}}_i}{\left\vert\sum_i \hat{\vec{r}}_i\right\vert}\,,
\end{equation}

\noindent and construct a local orthonormal frame $(\hat{\vec{e}}, \hat{\vec{f}}, \hat{\vec{n}})$ (east, north, center) via Gram--Schmidt, giving a tangent plane at the basin center. A naive $(\theta,\phi)$ or gnomonic projection onto this plane distorts distances from the center asymmetrically, biasing the fitted axis ratio and orientation. We instead use an azimuthal-equidistant projection centered on $\hat{\vec{n}}$ \citep{snyderMapProjectionsWorking1987}, which preserves the true angular distance from the center along every direction. For a rim pixel $\hat{\vec{r}}$, the angular distance from center is $c = \arccos(\hat{\vec{r}}\cdot\hat{\vec{n}})$, and its local Cartesian coordinates are

\begin{equation}
(x,y) = c\frac{\left(\hat{\vec{r}}\cdot\hat{\vec{e}},\hat{\vec{r}}\cdot\hat{\vec{f}}\right)}{\sqrt{(\hat{\vec{r}}\cdot\hat{\vec{e}})^2 + (\hat{\vec{r}}\cdot\hat{\vec{f}})^2}}\,.\label{eq:Crater_Center}
\end{equation}

\noindent Here $x$ and $y$ are in radians, so a distance in this plane equals the true angular distance on the sphere and converts to a physical length by multiplying by the body's radius $R$.

In this plane, we fit a general conic $Ax^2+Bxy+Cy^2+Dx+Ey+F=0$ to the projected rim points using the direct least-squares ellipse-specific method of \citet{fitzgibbonDirectLeastSquare1999}, as implemented in \texttt{scikit-image}'s \texttt{EllipseModel}. To remain robust against residual noisy rim pixels, the fit is embedded in a RANSAC scheme \citep{fischlerRandomSampleConsensus1981}: a small random subset of rim points proposes a candidate ellipse, points within a residual tolerance, set to \SI{5}{\percent} of the median angular rim distance from the basin center, are counted as inliers, and this is repeated over many trials, keeping the ellipse with the largest inlier support. This yields the ellipse center $(x_c,y_c)$, semi-major and semi-minor axes $a_{\rm rad}$ and $b_{\rm rad}$ (radians), and orientation angle $\psi$ in the tangent plane (we use $\psi$ to avoid confusion with the impact angle $\theta$).

Since the tangent-plane coordinates preserve true angular distance from the center exactly, the physical semi-major and semi-minor axes follow directly:

\begin{equation}
a = a_{\rm rad}\,R\,, \qquad b = b_{\rm rad}\,R\,,
\end{equation}

\noindent where $R$ is the planetary radius. Together with $\hat{\vec{n}}$ and $\psi$, this fully characterizes the basin outline as an ellipse on the sphere; for display, the fitted ellipse can be re-projected onto the sphere by inverting Equation~\eqref{eq:Crater_Center} and overlaid on the crustal-thickness map, as shown in Figure~\ref{fig:crater_map}.

All basin dimensions are measured at the final snapshot of each simulation (\SI{3.54}{} or \SI{1.77}{\hour} after impact; Section~\ref{sec:Initial_conditions}), well after the collapse of the transient cavity, whose gravitational timescale $\sqrt{D_{\rm eff}/g_{\rm M}}\lesssim\SI{15}{\minute}$ is short compared to either measurement time. For all simulations with snapshots at both \SI{1.77}{} and \SI{3.54}{\hour}, the measured $D_{\rm eff}$ is unchanged between the two times to within the precision of the ellipse fit; the basin diameter is therefore quasi-stationary well before the earlier measurement time. The diameters reported here thus characterize the immediate post-collapse basin, prior to any long-term viscous or isostatic adjustment (Section~\ref{sec:gravity_mantle}).

\section{Measuring the simulated gravity anomaly}\label{sec:gravity_method}
Here we describe how the gravity anomaly maps of Figure~\ref{fig:gravity_map}, discussed in Section~\ref{sec:gravity_mantle}, were computed from the simulation output, using the final snapshot of the 2-billion-particle simulation. The snapshot provides, for each particle, its mass and position $(x,y,z)$. Before computing the gravity field, we recenter the particle distribution on the center of mass of the post-impact body, using the same iterative procedure described in Appendix~\ref{sec:appendix:Measuring_Crater_Size} for the basin-size measurement.

To sample the gravity field produced by this recentered mass distribution, we generate a spherical shell of tracer points at a fixed radius of $\SI{0.4}{\Rearth}$ using a $N_{\rm side}=256$ HEALPix tessellation of the sphere \citep{gorskiHEALPixFrameworkHighResolution2005}, which distributes the tracers with equal area per pixel and thus avoids the latitude-dependent clustering of a naive longitude--latitude grid. Each tracer is assigned a negligible mass so that it acts as a passive test particle, sampling the field of the simulation particles without perturbing it. The tracer shell and the recentered simulation particles are combined into a single particle set and written out as a dark-matter (i.e., collisionless, gravity-only) particle file.

We then compute the gravitational acceleration at every particle, including the tracers, using the tree-based gravity solver of \texttt{pkdgrav3} \citep{potterPKDGRAV3TrillionParticle2017}, run in gravity-only mode on this combined particle set. \texttt{pkdgrav3} is well suited to this task as the same $N$-body gravity solver underlies the impact simulations themselves, ensuring a force calculation consistent with the simulation. The gravitational softening is set to zero in this post-processing step; at the separations between the tracer shell and the mass distribution, softening has no effect on the computed field.

To isolate the gravitational contribution of the deformed core (Section~\ref{sec:gravity_mantle}), we repeat the procedure with a restricted particle set: after recentering, only particles of core material (selected by their material identifier) are retained, converted to gravity-only particles, and combined with the same tracer shell. All subsequent steps are identical. Since the pre-impact core is spherically symmetric, its field on the tracer shell is uniform, so the core-only anomaly map requires no differencing against the pre-impact configuration: any structure in it reflects the impact-induced deformation of the core alone.

Finally, the resulting accelerations of each computation are read back in and restricted to the tracer shell. For each tracer we retain the magnitude of the acceleration vector, $\vert\vec{a}\vert$, and plot the difference from the mean of this quantity ($\vert\vec{a}\vert - \langle\vert\vec{a}\vert\rangle$) as a function of tracer longitude and latitude to produce the gravity anomaly map shown in Figure~\ref{fig:gravity_map}. Because the tracers sit at a fixed, common radius of $\SI{0.4}{\Rearth}$, spatial variations in $\vert\vec{a}\vert$ across the map directly reflect the underlying mass distribution of the recentered body, without the additional radius-dependent scaling that would arise if tracers were instead placed at, e.g., the (non-spherical) post-impact surface.


\bibliography{main}{}
\bibliographystyle{aasjournalv7}



\end{document}